\documentclass[journal,letter]{IEEEtran}

\usepackage[T1]{fontenc}
\usepackage{amsmath}
\usepackage{graphicx}
\usepackage{url}
\usepackage{xspace}
\usepackage{xcolor}
\usepackage{multirow}
\usepackage{array}
\usepackage{balance}
\usepackage{booktabs}
\usepackage{paralist}
\usepackage{threeparttable}
\usepackage{makecell}
\usepackage{algorithm}
\usepackage{algorithmic}
\usepackage[numbers]{natbib}
\usepackage{ragged2e} 
\usepackage{enumerate}

\usepackage{colortbl}
\usepackage{diagbox}
\usepackage{orcidlink}
\hypersetup{hidelinks}
\usepackage{amssymb}  
\usepackage{pifont}   
\usepackage{textcomp}

\usepackage{cleveref}

\colorlet{gray1}{gray!70}
\colorlet{gray2}{gray!25}
\colorlet{gray3}{gray!15}
\definecolor{darkgrey}{HTML}{434343}
\definecolor{lightgrey}{HTML}{A9A9A9}
\definecolor{silver}{HTML}{D3D3D3}
\definecolor{midgrey}{HTML}{808080}
\definecolor{white}{HTML}{FFFFFF}
\newcommand{\BAC}{BacAlarm\xspace}

\usepackage[most]{tcolorbox}
\newtcolorbox{LYbox}[2][]{text width=0.95\linewidth,fontupper=\normalsize,
  fonttitle=\bfseries\sffamily, colbacktitle=darkgrey,enhanced,
  attach boxed title to top left={yshift=-2mm,xshift=3mm},
  boxed title style={sharp corners},top=4pt,bottom=2pt,left=2pt,right=2pt,
  title=#2,colback=white,coltitle=white}

\def\BibTeX{{\rm B\kern.05em{\sc i\kern.025em b}\kern.08em
    T\kern.1667em\lower.7ex\hbox{E}\kern.125emX}}

\begin{document}

\title{\BAC: Mining and Simulating Composite API Traffic to Prevent Broken Access Control Violations}


\author{Yanjing Yang$^{\orcidlink{0009-0006-8789-4589}}$,
        He Zhang,
        Bohan Liu\IEEEauthorrefmark{1}$^{\orcidlink{0000-0002-0146-5411}}$,
        Jinwei Xu,
        Jinghao Hu,
        Liming Dong,
        Zhewen Mao,
        Dongxue Pan
\thanks{\IEEEcompsocthanksitem 
\IEEEcompsocthanksitem \emph{Corresponding author: Bohan Liu}
\IEEEcompsocthanksitem Yanjing Yang, He Zhang, Bohan Liu, Jinwei Xu, Jinghao Hu are with State Key Laboratory of Novel Software Technology, Software Institute, Nanjing University, Nanjing 210008, China (e-mail: yj\_yang@smail.nju.edu.cn; 
hezhang@nju.edu.cn; bohanliu@nju.edu.cn; 
jinwei\_xu@smail.nju.edu.cn; jinghao\_hu@smail.nju.edu.cn).
\IEEEcompsocthanksitem Liming Dong is with CSIRO Data 61, Australia (email: liming.dong@data61.csiro.au);
\IEEEcompsocthanksitem Zhewen Mao and Dongxue Pan are with Huawei Cloud Computing Technologies Co., Ltd., China (email: maozhewen@huawei.com, pandongxue@huawei.com)}}

\maketitle

\begin{abstract}

Broken Access Control (BAC) violations, which consistently rank among the top five security risks in the OWASP API Security Top 10, refer to unauthorized access attempts arising from BAC vulnerabilities, whose successful exploitation can impose significant risks on exposed application programming interfaces (APIs). In recent years, learning-based methods have demonstrated promising prospects in detecting various types of malicious activities. However, in real-network operation and maintenance scenarios, leveraging learning-based methods for BAC detection faces two critical challenges. Firstly, under the RESTful API design principles, most systems omit recording composite traffic for performance, and together with ethical and legal bans on directly testing real-world systems, this leads to a critical shortage of training data for detecting BAC violations. Secondly, common malicious behaviors such as SQL injection typically generate individual access traffic that is inherently anomalous. In contrast, BAC is usually composed of multiple correlated access requests that appear normal when examined in isolation. To tackle these problems, we introduce \BAC, an approach for establishing a BAC violation detection model by generating and utilizing API traffic data. The \BAC consists of an API Traffic Generator and a BAC Detector. 
The API Traffic Generator establishes a knowledge base from unlabeled historical logs and utilizes it to simulate composite traffic, providing training data that addresses the data scarcity caused by RESTful API design and ethical constraints.
The BAC Detector extracts sequential and static features from simulated data to train an ensemble-based model, providing augmented protection against BAC violations.
We evaluated \BAC on eight core function API-sets from two open-source systems with real-world CVE vulnerabilities. 
Experimental results show that \BAC outperforms current state-of-the-art invariant-based and learning-based methods with the $\text{F}_1$ and MCC improving by 21.2\% and 24.1\%. 
Both the Generator and the Detector incorporate algorithms we design to achieve strong effectiveness.
The Generator utilizes unsupervised API extraction with accuracy sufficient to replace manual endpoint identification, and its simulation method produces high-quality labeled composite traffic. The Detector achieves effective contextual modeling through the designed sequential features. Ablation studies further confirm that all parts are indispensable, as removing any component leads to performance degradation.


\end{abstract}

\section{Introduction}


Application Programming Interfaces (APIs) serve as the backbone of modern software systems. However, statistics reveal a striking gap between their importance and security practices: 95\% of industry professionals have encountered API security challenges, while only a mere 7.5\% of organizations implement dedicated measures to secure their APIs~\cite{salt2025api, sun2021research}. Broken Access Control (BAC) remains the most pressing concern among these risks. Attackers exploit BAC vulnerability to bypass authorization checks, obtain or alter resources belonging to other users. In severe instances, the allocation of administrative privileges may occur, potentially compromising entire systems. The 2023 OWASP API Security Top 10 report highlights its dominance, noting that four out of the top five threats are directly linked to BAC violations~\cite{owasp2023api}. BAC violations are malicious API requests from attackers who exploit BAC vulnerabilities. Successful BAC violations can cause serious harm, especially to sensitive APIs~\cite{sun2021research, almushiti2023investigation}. This disparity highlights the crucial need for improved access control mechanisms in API management~\cite{mosavi2023detecting, singh2025experiences}.

Static analysis~\cite{sun2011static, goseva2015capability, ghaleb2023achecker, lin2025paten} and dynamic testing~\cite{chehade2025403, liu2025bacscan} are able to detect certain BAC vulnerabilities.
Even though these methods are effective, they still fall short in real-world scenarios such as residual permission cookies after role changes or privilege propagation via shared resources.
To complement static and dynamic analysis, runtime protection based on API access request traffic\footnote{Hereafter, API access request traffic is referred to as API traffic.} leverages dynamic execution logs to detect abnormal behaviors and block or delay malicious executions.

Existing runtime protections are mainly divided into two categories, each with its own limitations. The first is invariant-based methods~\cite{marinescu2017ivd, linblock, karimi2021automatic}. These methods entirely rely on historical traffic to identify invariants (\texttt{$<role, resource, action>$}), thereby setting rules and establishing whitelists. As a result, they impose high implementation and maintenance costs.
Another more promising category is learning-based methods~\cite{meng2019loganomaly, Liu-fur, wang2023research, anas2024bacad}, which focus on single API traffic and thus fail to capture dependencies across multiple requests.
For instance, CVE-2022-31133\footnote{CVE-2022-31133.\url{https://www.cvedetails.com/cve/CVE-2022-31133/}} demonstrates a BAC vulnerability that can be exploited by composite API traffic. It allows cookie theft to impersonate users in API traffic, enabling BAC violations like deleting comments. Each request individually appears valid, making detection difficult without a composite API traffic context. This implies that existing learning-based methods targeting single traffic are unable to block such attacks due to the lack of context-sensitive awareness.

To accurately capture such a context-sensitive pattern through learning-based methods requires overcoming two challenges. The first challenge lies in the unavailability of labeled composite traffic data. This is primarily because ethical and legal restrictions (e.g., GDPR~\cite{GDPR}) prevent collecting such data from real users~\cite{chehade2025403}, while the efficiency-oriented design of RESTful APIs inherently hinders tracking correlations across sequential accesses~\cite{petrillo2016rest, bogner2023restful}. 
The second challenge lies in the lack of effective methods for extracting features from composite API traffic. Existing approaches mainly focus on single-API and thus fail to capture the cross-request dependencies and behavioral patterns that are essential for identifying BAC violations formed by composite API traffic.
As a result, current runtime protections are incapable of capturing the context-sensitive pattern necessary for detecting BAC violations.

This study aims to tackle the challenges in the learning-based detection of BAC violations. We present \BAC, an anomaly detection approach comprising two core components: an \textbf{API traffic generator} and a \textbf{BAC detector}. The API traffic generator first parses API traffic logs to extract structured information, constructing a comprehensive API knowledge base that captures the normal behavior patterns of the targeting system. Leveraging the API knowledge through the Retrieval-Augmented Generation (RAG) technique, it prompts a Large Language Model (LLM) to produce simulated composite API access requests, including both benign and malicious scenarios. The BAC detector further extracts bespoke sequential features from the simulated traffic and then employs the designed ensemble learning model to train a robust anomaly detection model.

To evaluate the effectiveness of \BAC, we derive eight core function API-sets from two open-source systems that have been reported to contain BAC vulnerabilities in their respective CVE disclosures. Experimental results show that \BAC outperforms current State-Of-The-Art (SOTA) invariant-based and learning-based methods, achieving improvements of 21.2\% in $\text{F}_1$ and 24.1\% in MCC. Across the three key phases of our approach, we further conduct detailed comparisons against representative candidates. In the API traffic generator, our unsupervised API extraction method outperforms existing baselines by 3.55\% in precision and 9.7\% in recall, and our data simulation approach achieves the highest fidelity with the lowest cost while yielding the best BAC violation detection performance. In constructing the BAC detector, the designed API-syntax features and composite-traffic transfer-entropy features rank among the most effective. Ablation studies confirm that removing API information, the hallucination-elimination module, or our sequence features leads to significant performance degradation.
These findings highlight that simulated data using \BAC can effectively train detection models and underscore the critical importance of detecting BAC violations by composite traffic.

The contributions of this article are summarized as follows.
\begin{itemize}

    \item \textbf{Method:} We propose an ethical API traffic generation simulation method for composite API traffic in BAC violation.

    \item \textbf{Model:} We develop a detection model specifically targeting composite-traffic-based BAC violation.
    
    \item \textbf{Data:} We construct two test datasets covering 8 functional API sets from 2 real applications with known BAC vulnerabilities, along with simulated API traffic datasets.
    
\end{itemize}

The remainder of this article is organized as follows. ~\Cref {sec: relatedwork} discusses the related work. ~\Cref{sec: Motivation} presents the motivation for designing \BAC. \Cref{sec: approach} presents the proposed \BAC. ~\Cref{sec: experiment} presents the experimental designs and results analysis on evaluating \BAC. ~\Cref{sec: discussion} discusses the effectiveness of sequence features, the validity of simulated data in BAC violation detection, and the impact of integrating \BAC with existing built-in defense mechanisms. Finally, we present the conclusion and the future work in ~\Cref{sec: conclusion}.

\section{Related Work}
\label{sec: relatedwork}


Recent research on BAC vulnerability detection spans both development and runtime stages. Early approaches apply static analysis~\cite{sun2011static, ghaleb2023achecker, Monshizadeh2014MACEDP} or automated testing~\cite{chehade2025403, liu2025bacscan, kushnir2021automated} to identify permission inconsistencies by analyzing access logic or replaying role-specific requests. However, reliance on source code or full interface coverage limits real-world applicability and precludes real-time protection. To address these issues, recent work shifts toward runtime detection methods based on real API traffic, which fall into two main categories: invariant-based modeling and anomaly detection.



    


Invariant-based modeling extracts stable access patterns from historical traffic and flags violations at runtime. 
BLOCK~\cite{linblock} treats applications as stateless systems and mines request–response transitions to infer expected access flows, detecting inconsistencies that indicate access control violations.
DetLogic~\cite{DEEPA201889} builds finite-state machines to model allowed state transitions and identify logic flaws that break access control constraints.
IVD~\cite{marinescu2017ivd} extracts user–resource invariants from database access logs to establish normal access relationships, and flags any requests violating these invariants as potential access control violations.
Although these methods perform well within their settings, they often require source code, assume stateless sessions, or depend on fixed environments, which limit their applicability and scalability in modern, highly dynamic web systems.

Learning-based anomaly detection model captures API behavioral deviations in runtime traffic.
DoubleGuard~\cite{le2011doubleguard} enforces per-session consistency via container isolation, but its reliance on instrumentation limits deployment to controlled environments.
SENTINEL~\cite{li2012sentinel} builds access behavior models from SQL logs to detect unauthorized database operations, but its database-centric design limits applicability to traffic-level access control violations.
LogAnomaly~\cite{meng2019loganomaly} proposes a unified anomaly detection framework that models log streams as semantic sequences, enabling simultaneous detection of both sequential and quantitative anomalies in highly dynamic, large-scale systems.
BRM~\cite{wang2023research} applies geometric feature metrics to quantify deviations in traffic behavior, but without semantic understanding, it fails to distinguish benign workflow variations from true violations.
BACAD~\cite{anas2024bacad} classifies API calls through proxy-based role models to identify cross-role access attempts, yet lacks the capacity to capture sequential dependencies across composite traffic.
\begin{table}[hbpt]
\scriptsize
\centering
\begin{threeparttable}
\caption{Comparison of BAC detection approaches across development and runtime stages.}
\renewcommand{\arraystretch}{1.1}
\setlength{\tabcolsep}{2pt}
\begin{tabular}{%
p{2.3cm}  
p{0.9cm}  
p{1.8cm}  
p{1.8cm}  
r          
}
\toprule
\textbf{Method} & \textbf{Stage} & \textbf{Context Sensitive} & \textbf{Ethical Constraint$^\dagger$} & \textbf{Methodology} \\
\midrule
A-CHECKER~\cite{ghaleb2023achecker} & Dev. & \ding{55} & \ding{55} & Rule \\
MACEDP~\cite{Monshizadeh2014MACEDP} & Dev. & \ding{55} & \ding{55} & Rule \\
Active~\cite{kang2021active} & Dev. & \ding{55} & \ding{55} & Rule \\
Paten~\cite{lin2025paten} & Dev. & \ding{55} & \ding{55} & Rule \\

API-MCTree~\cite{tian2018automatically} & Test. & \ding{55} & \ding{55} & Rule \\
VSF~\cite{chehade2025403} & Test. & \ding{55} & \checkmark & Fuzz \\
BACscan~\cite{liu2025bacscan} & Test. & \checkmark & \ding{55} & Fuzz \\
IVD~\cite{marinescu2017ivd} & Run. & \ding{55} & \ding{55} & Invariant \\
BLOCK~\cite{linblock} & Run. & \ding{55} & \ding{55} & Invariant \\
DetLogic~\cite{DEEPA201889} & Run. & \ding{55} & \ding{55} & Invariant \\
DoubleGuard~\cite{le2011doubleguard} & Run. & \ding{55} & \ding{55} & Learning \\
SENTINEL~\cite{li2012sentinel} & Run. & \ding{55} & \ding{55} & Learning \\
LogAnomaly~\cite{meng2019loganomaly} & Run. & \checkmark & \ding{55} & Learning \\
BRM~\cite{wang2023research} & Run. & \ding{55} & \ding{55} & Learning \\
BACAD~\cite{anas2024bacad} & Run. & \ding{55} & \ding{55} & Learning  \\
\BAC(Ours) & Run. & \checkmark & \checkmark & Learning  \\
\bottomrule
\end{tabular}
\begin{tablenotes}
\footnotesize
\item[\dag] \textbf{Ethical Constraint} indicates whether the approach can operate without relying on real user data, thereby ensuring compliance with privacy and ethical regulations during deployment.
\end{tablenotes}
\label{tab:bac_comparison}
\end{threeparttable}
\end{table}

Learning-based methods, particularly deep learning models, have proven effective at capturing the complex, cross-request behavioral patterns associated with BAC violations~\cite{meng2019loganomaly, anas2024bacad}. These models can automatically learn sequential dependencies and identify intricate patterns from historical data, offering better scalability and adaptability than traditional rule-based or invariant approaches, which rely on predefined rules and struggle with new request types or evolving traffic patterns. As summarized in~\Cref{tab:bac_comparison}, although these methods effectively detect diverse behavioral anomalies, they typically adapt to fixed environments, require additional specific tools, and focus on a single request, limiting their scalability and generalization to dynamic multi-session composite API traffic BAC violations.


\begin{figure}[htbp]
    \centering
    \begin{minipage}{1.0\linewidth}
        \centering
        \includegraphics[width=\linewidth]{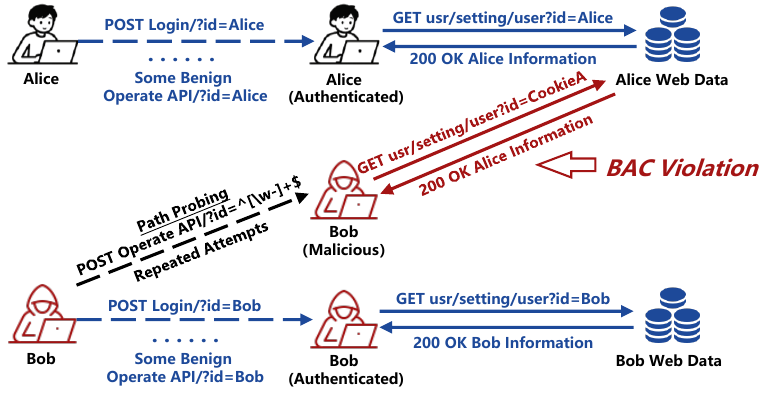}    
        \footnotesize (a) Modeling the Attack Scenario of BAC Violations
    \end{minipage}

    \begin{minipage}{0.98\linewidth}
        \centering
        \includegraphics[width=\linewidth]{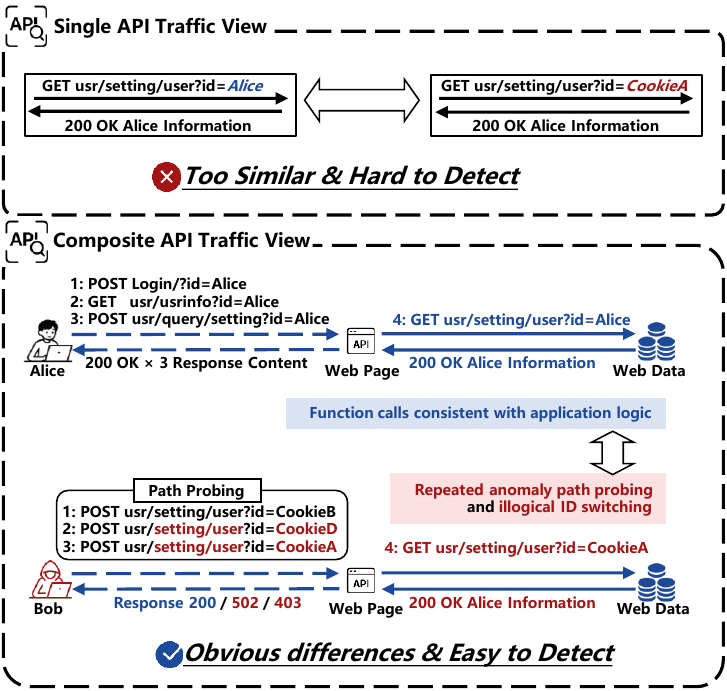}
        \footnotesize (b) Detection Differences Across Single and Composite API Traffic Views
    \end{minipage}

    \caption{Motivating the Use of Composite Traffic Analysis for Reliable BAC Violation Detection}
    \label{fig: Motivation}
    \vspace{-2.5ex}
\end{figure}

\section{Motivation}
\label{sec: Motivation}

Unlike other common attacks such as injection attacks, DDoS, etc., a striking feature of BAC violations is that each piece of traffic generated by implementing or attempting unauthorized access requests may appear entirely normal, while the behavior composed of a series of access requests initiated during unauthorized attack attempts is abnormal. This leads to the fact that although learning-based methods are widely used in web security, their successful applications in BAC violation detection remain extremely limited.

For example, malicious user Bob can exploit CVE-2022-31133 to obtain several residual session cookies. With these cookies in hand, Bob begins probing the application by issuing repeated trial API requests, attempting to discover whether any of them can be misused to access another user’s resources. As shown in \Cref{fig: Motivation}(a), Bob performs path probing operations and repeatedly tests API endpoints that may lead to BAC violations during this exploration phase. Once Bob identifies a feasible opportunity, Bob successfully accesses Alice’s sensitive information by sending an unauthorized request using the stolen cookie. After completing the unauthorized access, he then issues several benign operations with his legitimate CookieB to blend into benign traffic patterns and obscure the preceding BAC violations.
It illustrates how a malicious user can exploit a BAC vulnerability to perpetrate cross-privilege-boundary violations by abusing authentication tokens. The attack unfolds through a multi-step composite API traffic sequence, where seemingly legitimate requests are weaponized to execute malicious operations across diverse user contexts.

As shown in~\Cref{fig: Motivation}(b), in the single API traffic analysis, each request appears entirely benign. For example, a \texttt{GET usr/setting/user?id=CookieA} returning \texttt{200 OK} looks almost indistinguishable from a normal \texttt{GET usr/setting/user?id=Alice} that also returns \texttt{200 OK}. Because both the API traffic syntax and the response status are similar, detecting BAC violation using a single traffic is fundamentally unable to distinguish between malicious and benign accesses. The abnormal switching between authentication tokens, the inconsistent user identities across adjacent requests, and the unusual temporal ordering of operations collectively expose patterns that are invisible at the single API traffic level. Only by analyzing these cross-request contextual relationships of composite API traffic can BAC violations be effectively detected.

Insights from related work indicate that effectively capturing such cross-request contextual relationships requires more powerful learning-based approaches capable of modeling sequential and composite behaviors. However, any learning-based BAC detector must be trained and evaluated under strict ethical constraints (e.g., using only researcher-controlled accounts and experiments that avoid exposure or use of third-party user data~\cite{chehade2025403}). Motivated by the related work in~\Cref{tab:bac_comparison} and the protection requirements observed in real industrial environments, this work investigates a learning-based runtime approach that, under ethical restrictions, models composite API-traffic contexts to enable the detection of cross-session composite BAC violations in real-world web applications.

\section{Methodology}
\label{sec: approach}

To construct a context-sensitive learning-based BAC violation detector, we first discuss the challenges of learning-based methods in BAC detection in ~\Cref{Met: challenge}. Then, we introduce two core components of \BAC: the API Traffic Generator and the BAC Detector, which are essential for scalable and effective BAC violation detection in ~\Cref{Met: generator} and ~\Cref{Met: detector}.

\begin{figure*}[htbp]
\centering
\includegraphics[width=1.0\textwidth]{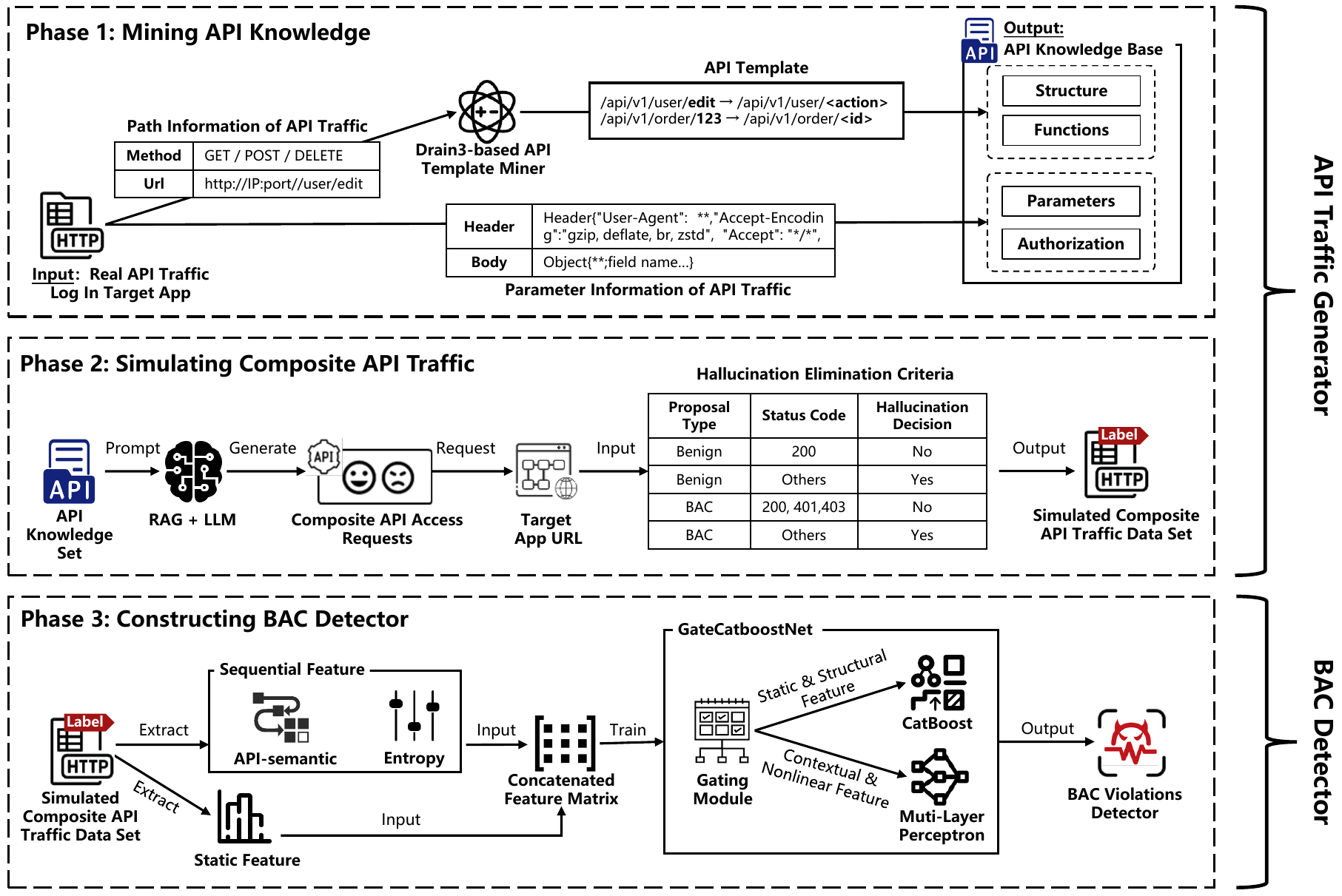} 
\vspace{-4.0ex}

    \caption{The Workflow of \BAC}
\label{fig: approach}
\vspace{-2.0ex}
\end{figure*}

\subsection{Challenges of Constructing Learning-based Detector}
\label{Met: challenge}



In this research, we focus on detecting BAC violations in open API web applications. Despite the promise of learning-based methods~\cite{anas2024bacad, wang2023research, FUR-API} for detecting such violations, there are two key challenges in applying these approaches.

\textbf{Challenge 1: Lack of ethical collection of labeled composite traffic.}
Although many existing datasets provide the labeled single traffic, they lack annotations for composite traffic associated with malicious user behaviors, so they do not capture the contextual relationship of more complex BAC violations. Ethically labeled composite traffic is challenging due to privacy concerns, particularly under regulations like GDPR, which prohibit unauthorized access to sensitive user data. Additionally, these violations are typically low-frequency and covert, making them difficult to capture in real-world data. Simulated traffic generation presents a viable solution to this issue, as it allows the creation of labeled composite traffic without compromising user privacy. By generating API traffic that simulates real-world user behavior, researchers can create more comprehensive datasets for training and testing BAC detectors.

\textbf{Challenge 2: The Lack of Effective Context-Aware Learning Methods.}
For effective BAC violation detection, learning methods must be context-aware, meaning they need to understand the user behavior across different requests. While the existing method~\cite{chehade2025403} has shown success in dynamic environments under ethical constraints, it may not fully address the detection of complicated BAC violations caused by composite API traffic.

The core assumption behind \BAC is that a user's behavior in a web application follows a predictable baseline. If a user's API usage deviates significantly from this baseline, it may indicate a BAC violation. 
As shown in ~\figurename~\ref{fig: approach}, \BAC consists of two core components: the API Traffic Generator and the BAC Detector. The API Traffic Generator is designed to tackle Challenge 1 by extracting an API knowledge set from real traffic using a Drain3~\cite{he2017drain} to capture API endpoint information. With this knowledge, an LLM-based agent generates labeled composite API traffic, helping to address the challenge of ethically obtaining labeled composite API traffic for training. The BAC Detector is designed to address Challenge 2 by capturing context-sensitive complex behavioral patterns. We propose sequential features to assist the original static features, allowing for more accurate detection of BAC violations that span multiple composite API traffic.


\subsection{API Traffic Generator}
\label{Met: generator}

To address the lack of labeled composite API traffic data labels (Challenge 1: Lack of Labelled Data), we independently register new users and simulate API traffic data with associated labels. The process of generating simulated data by the API Traffic Generator consists of two phases: (1) mining API knowledge and (2) simulating composite API traffic.

\textbf{Mining API Knowledge Phase:} A structured API knowledge base is constructed from raw traffic logs collected during application initialization and automated interface crawling. To extract meaningful API activities from noisy traffic, we design a Drain3-based API discovery algorithm that filters and parses request traces to generate generalized API representations. The algorithm first processes collected logs to isolate relevant API calls and then applies template mining to extract structural patterns. The algorithm captures stable request formats while masking variable fields (e.g., IDs or tokens), producing standardized templates for each endpoint.

\begin{figure}[htbp]
\centering
\includegraphics[width=0.85\linewidth]{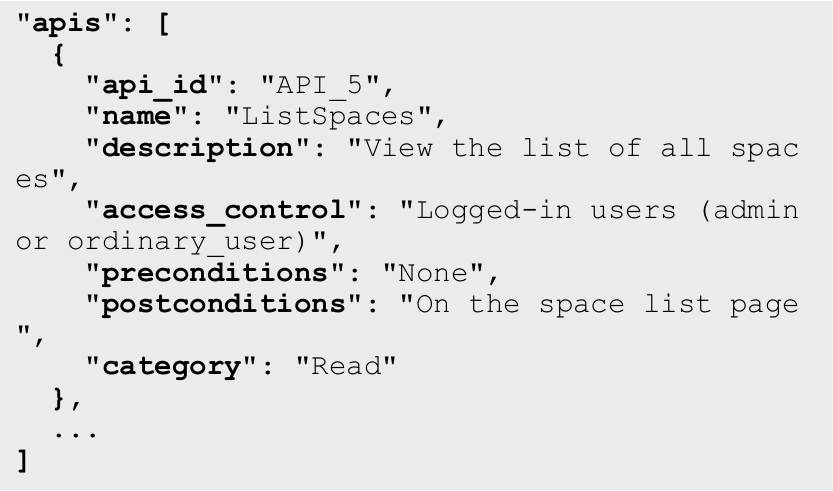}
\caption{Example of a Structured API Knowledge Item}
\label{fig:knowledge_item}

\end{figure}

These templates serve as the basis for extracting four key types of information: (1) structure (i.e., HTTP methods, normalized URL formats), (2) initial functional semantics, (3) allowed parameter values, and (4) authorization indicators. These entries integrate structural patterns with functional insights, culminating in a comprehensive knowledge base for downstream applications. A representative example is shown in~\figurename~\ref{fig:knowledge_item}.

\begin{figure}[htbp]
\centering
\includegraphics[width=0.95\linewidth]{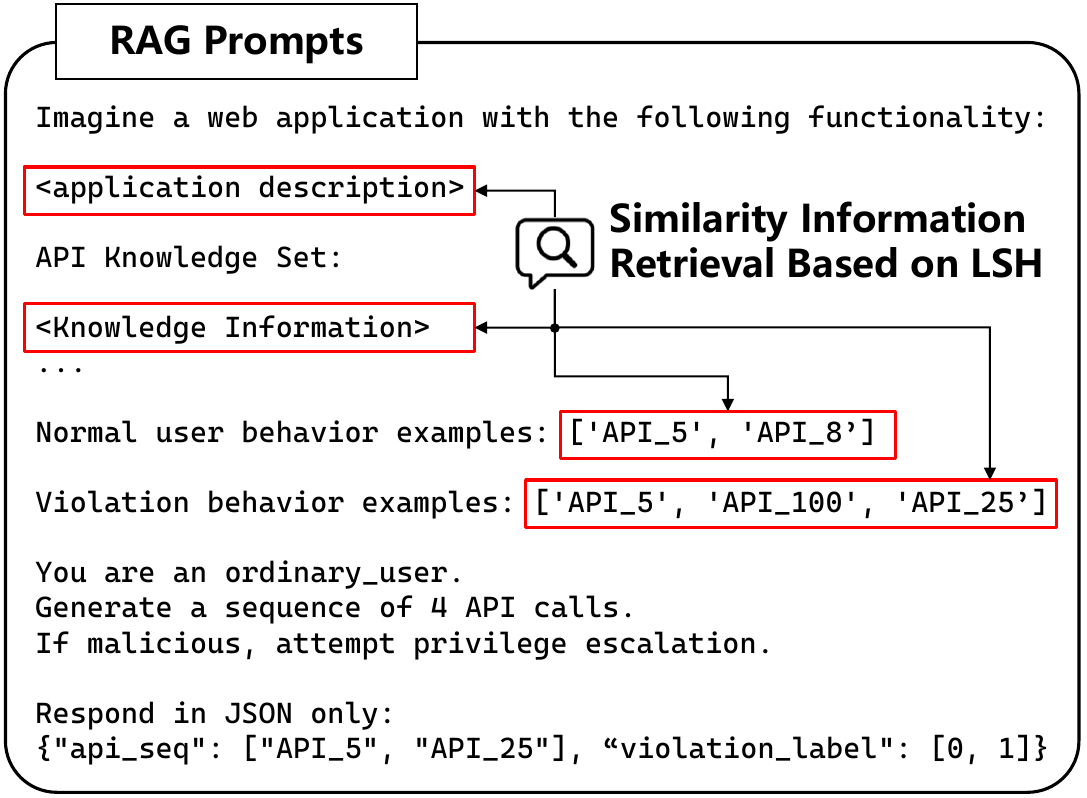}
\caption{RAG Prompts in \BAC}
\label{fig: RAG}
\vspace{-3.0ex}
\end{figure}

\textbf{Simulating Composite API Traffic Phase:} The objective of this phase is to generate valid API traffic, encompassing both legitimate traffic and malicious traffic intended for privilege escalation, targeting a specified entry URL.

To be specific, each data generation cycle randomly assigns the LLM one of two roles (50\% probability each): generating either legitimate or malicious API access patterns. 
The workflow proceeds as follows: 1) LLM Generates Behavior Description: The LLM first crafts a natural language description of the target access behavior guided by  (e.g., "update user profile" or "exfiltrate database credentials"). 2) RAG Retrieves Relevant APIs:
The behavior description is passed to a RAG module, which uses Locality-Sensitive Hashing (LSH) to search the API knowledge base and retrieve APIs that are semantically related to the described behavior (e.g., a mention of ``user profile'' leads RAG to retrieve \texttt{/api/users/\{id\}/profile}). As illustrated in \figurename~\ref{fig: RAG}, RAG leverages LSH-based similarity search to provide behavior-relevant API endpoints that support the construction of RAG prompts. 
3) Prompt Engineering: RAG constructs a context-dependent prompt using the retrieved APIs (e.g., "Generate a sequence accessing /login, /profile, and /update\_email to simulate a normal user session"). 4) LLM Generates Access Sequence: The LLM processes the prompt and outputs a sequence of API access requests (potentially spanning multiple interdependent requests, such as authentication followed by data retrieval). 5) Automated Execution: The generated sequence is executed against the target URLs via automated HTTP requests. Each request logs detailed traffic metadata (headers, payloads, timestamps) and captures the HTTP status code. 6) Quality Filtering: The collected traffic is filtered using the Hallucination Elimination Criteria to discard invalid or inconsistent sequences. Benign requests are retained only if they return \texttt{200}, while BAC violation requests are accepted if they return \texttt{200}, \texttt{401}, or \texttt{403}. Requests triggering error codes (e.g., \texttt{400} or \texttt{404}) or contradicting their declared intent are tagged as hallucinations and discarded. The filtered results form a simulated composite API traffic dataset, serving as training data for the BAC Detector.

As illustrated in~\figurename~\ref{fig: RAG}, we employ Retrieval-Augmented Generation (RAG) powered by Locality-Sensitive Hashing (LSH)~\cite{he2025evaluating} to generate prompt texts derived from the entry URL and the API knowledge base, which are subsequently input into an LLM for request generation.

\subsection{BAC Detector}
\label{Met: detector}
To address the limited effectiveness of existing context-sensitive learning methods (Challenge~2: Lack of Context-Sensitive Feature), we collect a set of existing statistical features and further propose two types of sequential features: (1) API-semantic features and (2) Entropy-based features to capture anomalous contextual patterns within composite API traffic more effectively. In this study, we use the API Traffic Generator to generate 500 composite simulated API traffic sets per application for training purposes.

\textbf{Static features} are collected through a systematic literature retrieval and screening process. Using a predefined search string for BAC violation detection (e.g., “API anomaly detection”, “access control violation”, “behavioral features”, and “API privilege misuse”), we searched across DBLP, IEEE~Xplore, ACM Digital Library, and Google Scholar. After reviewing the abstracts, introductions, and methodologies of the filtered papers, we selected the following representative studies~\cite{ goseva2012classification, jeeva2016intelligent, kang2021active, christodorescu2022privacy, aharon2025classification}. From these works, we collected and synthesized 13 static features, listed in~\Cref{tab: featureBaselineList}, which characterize anomalous API traffic behaviors such as unusual access patterns, oversized request parameters, and repetitive probing attempts.

\begin{table}[htbp]
\centering
\scriptsize
\caption{Features for API Traffic Sequences}
\label{tab: featureBaselineList}
\renewcommand{\arraystretch}{1.3}
\begin{tabular}{@{}p{0.25\columnwidth}|p{0.48\columnwidth}|p{0.15\columnwidth}@{}}
\hline
\textbf{Feature Name} & \textbf{Description and Calculation Method} & \textbf{Ref} \\
\hline

UniquePathsCount & Number of distinct API paths in requests. & \cite{BACTypes-almushiti2023investigation,Liu-fur}\\ 
TotalPathsCount & $N$, total number of requests. & \cite{Liu-fur,anas2024bacad}\\ 
UniqueParamsCount & Number of unique query parameter keys used across all requests. &\cite{BACTypes-almushiti2023investigation, anas2024bacad, Liu-fur} \\ 
TotalParamsCount & Total number of query parameters (counting all occurrences). &\cite{anas2024bacad, zhang2019robust} \\ 
ConsecutiveRepeats & Number of times the same path appears in adjacent requests. & \cite{10456177, BACTypes-almushiti2023investigation, anas2024bacad}\\ 
AvgPathLength & Average number of characters in each request path. & \cite{BACTypes-almushiti2023investigation,ghaleb2023achecker, 10456177} \\
StdPathLength & Standard deviation of path lengths (in characters). & \cite{BACTypes-almushiti2023investigation,anas2024bacad}\\ 
AvgParamCount & Average number of query parameters per request. & \cite{BACTypes-almushiti2023investigation, zhang2019robust, ghaleb2023achecker}\\ 
StdParamCount & Standard deviation of query parameter counts per request. & \cite{zhang2019robust, BACTypes-almushiti2023investigation,anas2024bacad}\\ 
AvgPathDepth & Average number of slashes “/” in each request path. &\cite{10456177, BACTypes-almushiti2023investigation, anas2024bacad} \\ 
StdPathDepth & Standard deviation of path depths. & \cite{meng2019loganomaly, 10456177, anas2024bacad}\\
UniquenessRatio & Ratio of unique paths to total requests. & \\ 
StatusCodeDiversity & Number of distinct status codes observed in the requests. & \cite{BACTypes-almushiti2023investigation,anas2024bacad}\\

\hline
\end{tabular}
\end{table}

\textbf{Sequential features} are newly designed to capture the temporal and procedural interdependencies within composite API traffic, comprising two categories: (1) API-semantic features that encode the functional intent of API calls, and (2) entropy-based features that quantify the uncertainty and structural variations across request sequences.

\textbf{(1) API-syntax features.}
Well-behaved applications exhibit a stable API-calling syntax, reflected in the consistent ordering and dependency structure among API events. Malicious or
abnormal behaviors typically break this implicit syntax. To capture these sequential semantics, we treat API calls as discrete tokens and learn their normal
transition patterns through auto-regressive sequence modeling.

Given a sequence of API events $(E_1,\dots,E_T)$, the model uses a
causally-masked Transformer encoder to estimate the conditional distribution of
each next event, ensuring that the representation at position $t$ depends only
on the past events $(E_1,\dots,E_{t-1})$. The model is trained in an
auto-regressive manner to maximize the likelihood of the observed transitions:
\begin{equation}
\mathcal{L}(\theta)
= \sum_{t=2}^{T} \log p_\theta(E_t \mid E_1, \dots, E_{t-1}),
\label{eq:objective}
\end{equation}
allowing it to capture the temporal regularities that characterize benign API
traffic and to quantify deviations from these learned syntax.


At inference stage, the model evaluates a user sequence by computing the negative log-likelihood of each observed transition. For each event $E_t$, we define its deviation score in \Cref{eq:dev-score}:
\begin{equation}
\label{eq:dev-score}
S_t = -\log P(E_t \mid E_{1:t-1}),
\end{equation}
where larger values indicate more improbable transitions. To reflect that user intent becomes clearer in later stages of a session, we apply an exponential positional weighting shown in \Cref{eq:pos-weighting}:
\begin{equation}
\label{eq:pos-weighting}
S = \frac{\sum_{t=1}^{T} S_t \cdot \exp(t/T)}{\sum_{t=1}^{T} \exp(t/T)},
\end{equation}
where $T$ denotes the sequence length. The resulting scalar $S$ serves as the sequence-level anomaly feature and quantifies how strongly a user's API call behavior deviates from established benign patterns.

\textbf{(2) Entropy-based features} quantify the variability and structural heterogeneity within a user's API request sequence. For any categorical request attribute $x$ (e.g., HTTP method, response status code, or URL path), we compute its Shannon entropy as defined in \Cref{eq:entropy}:
\begin{equation}
\label{eq:entropy}
H_x = - \sum_i p_{x,i} \log_2 p_{x,i},
\end{equation}
where $p_{x,i}$ denotes the empirical probability of category $i$ for attribute $x$. A higher $H_x$ indicates greater diversity in the user's behavior along that attribute. To characterize structural complexity in the evolution of an attribute sequence, we additionally compute its first-order transition entropy, shown in \Cref{eq:trans-entropy}. Given a sequence $\{x_1, x_2, \dots, x_T\}$, we enumerate adjacent transitions $(x_t, x_{t+1})$ and compute:
\begin{equation}
\label{eq:trans-entropy}
H_{\text{Tran(x)}} = - \sum_{i,j} p_{x,ij} \log_2 p_{x,ij},
\end{equation}
where $p_{x,ij}$ denotes the empirical probability of transitioning from state $i$ to $j$ for attribute $x$. A higher $H_{\text{}}$ indicates more irregular or less predictable navigation patterns in that attribute. Based on \Cref{eq:entropy,eq:trans-entropy}, we include: the entropy of HTTP method diversity ($H_{\text{method}}$) and its transition complexity ($H_{\text{Trans(method)}}$); the entropy of status code focus variability ($H_{\text{status(200, 403, and 502, etc.)}}$), the entropy of status transitions ($H_{\text{Trans(status)}}$), and an aggregated uncertainty entropy ($H_{\text{SUM(status)}}$); and finally, the entropy capturing URL path ($H_{\text{path}}$) along with ttransition complexity ($H_{\text{Trans(path)}}$).

\textbf{Detector Construction} integrates the statistical features and the two categories of sequential features through a hybrid classification architecture, \textit{GatedCatBoostNet}, which unifies these heterogeneous representations in a single model.
The entropy-based statistical features and structural attributes align well with tree-based models, which effectively handle tabular and mixed-type inputs and remain robust under limited data.
In contrast, the API-syntax features capture sequential dependencies and nonlinear behavioral patterns arising from temporal API-call dynamics, making them better suited for a lightweight neural network that can model contextual transitions and detect deviations from normal API-syntax. To leverage the strengths of both modeling approaches, the architecture combines a CatBoost expert with a compact multi-layer perceptron (MLP). A gating module then learns how to fuse the two predictors by assigning a weight to each expert based on the input feature vector. The final prediction for an input $x$ is computed as shown in \Cref{eq:gatedcatboostnet}: 
\begin{equation}
\label{eq:gatedcatboostnet}
\mathbf{p}(x) = \underbrace{g_{\text{CB}}(x)}_{\text{gate weight}} \cdot \underbrace{f_{\text{CB}}(x)}_{\text{CatBoost expert}} \;+\; \underbrace{g_{\text{MLP}}(x)}_{\text{gate weight}} \cdot \underbrace{f_{\text{MLP}}(x)}_{\text{MLP expert}},
\end{equation}
where the gating weights satisfy $g_{\text{CB}}(x)+g_{\text{MLP}}(x)=1$ and are produced by a softmax layer. This design allows GatedCatBoostNet to combine the interpretability and the ability of gradient-boosted trees to handle structured feature data with the flexibility of neural representations, enabling the classifier to adapt to each sample and capture complex behavioral patterns more effectively.

\section{Experimental Design}
\label{sec: experiment}
\subsection{Research Questions}

To comprehensively evaluate the effectiveness of \BAC, we design the following research questions:

\begin{itemize}
    \item \textbf{RQ1:} How effective is \BAC in detecting BAC violations compared to existing detection methods?

    \item \textbf{RQ2:} How accurately can \BAC extract API-level knowledge automatedly from unlabeled API traffic?

    \item \textbf{RQ3:} How effective is the LLM-based Agent simulation module in BAC violation detection?

    \item \textbf{RQ4:} How effective are the sequential features learned by \BAC for detecting BAC violations?

    \item \textbf{RQ5:} How does each core component of \BAC contribute to the overall system performance?
\end{itemize}

The five research questions (RQ1–RQ5) collectively provide a holistic evaluation of \BAC. RQ1 serves as an overall comparison, assessing the detection performance of \BAC relative to existing state-of-the-art methods. RQ2 and RQ3 evaluate the two core components of the API Traffic Generator: RQ2 examines its ability to extract API-level knowledge from unlabeled API traffic (Phase 1), while RQ3 assesses the effectiveness of its LLM-based agent simulation module (Phase 2). RQ4 focuses on the BAC Detector, evaluating the effectiveness of the designed sequential feature in BAC violation detection (Phase 3). Finally, RQ5 conducts an ablation study to isolate and measure the contribution of each individual component to the overall system performance. This sequence of questions ensures a thorough validation of both the Generator and the Detector as well as the entire \BAC framework.

\subsection{APIs for Experiments}
\label{sec:dataset}

We selected two representative open-source web applications, Humhub~\cite{HumHub} and Memos~\cite{Memos}, based on their popularity (over 5,000 stars on GitHub), active maintenance, browser-accessible interfaces, and built-in authentication and access control mechanisms. In particular, both systems exhibit documented and reproducible BAC vulnerabilities, such as CVE-2022-31133 and CVE-2022-4690, which serve as a reliable foundation for constructing realistic exploit scenarios.

\begin{table}[htbp]
\vspace{-2.0ex}
\scriptsize
\centering
\caption{Target Applications Overview}
\label{tab:target_apps}
\begin{tabular}{@{}>{\raggedright\arraybackslash}p{0.13\columnwidth}|
                >{\raggedright\arraybackslash}p{0.4\columnwidth}|
                >{\raggedright\arraybackslash}p{0.20\columnwidth}|
                >{\raggedright\arraybackslash}p{0.06\columnwidth}@{}}
\hline
\textbf{Application} & \textbf{Description} & \textbf{Security Issue} & \textbf{Star} \\
\hline
Humhub & A platform supporting user and content management, with RBAC (Role-Based Access Control) and ABAC (Attribute-Based Access Control). & CVE-2022-31133: Script in space name runs on join, enabling cookie theft. & 6.4k \\
\hline
Memos & A note-taking application that manages both file and user information, incorporating RBAC+ABAC to enforce fine-grained access control policies & CVE-2022-4690: JS in SVG triggered on view, allowing XSS attack. & 38.3k \\
\hline
\end{tabular}
\vspace{-2.0ex}
\end{table}

\begin{table}[htbp]
    \caption{Basic Information of Test API-sets}
    \label{tab:API-setOverview}
    \centering
    \scriptsize
    \renewcommand\arraystretch{1.2}
    \setlength{\tabcolsep}{0.8mm}
    \begin{tabular}{c|p{3.8cm}|c|c|c}
        \hline
        \textbf{API-set} & \textbf{Function} & \textbf{\#Requests} & \textbf{\#Violations} & \textbf{\#Exploits} \\
        \hline
        API-set1  & Space and comment actions with role-based control     &  6776  & 1134 & 589 \\
        API-set2  & Space, user, and notification management; admin-only module settings & 5298 & 1642 & 847 \\
        API-set3  &Space interaction and comment management with role-based permissions & 6407 & 1728 & 878 \\
        API-set4  & User account management and social interactions with scoped access & 6122 & 1837 & 938 \\
        API-set5  & Space access and invitation, notification viewing, and restricted module management & 6145 & 1741 & 930 \\
        API-set6  & Memo creation and browsing, personal settings control, and restricted system configuration & 6489 & 2424 & 1504 \\
        API-set7  & Memo viewing and labeling, resource access, user creation, and settings management under RBAC control & 3458 & 1745 & 627 \\
        API-set8 & Memo viewing by creator, user-specific settings management, and admin-only user creation under RBAC control  & 5070 & 1226 & 726 \\
        \hline
    \end{tabular}
\end{table}

We manually constructed our dataset due to the absence of publicly available datasets for BAC detection. We selected two popular open-source projects with documented access control vulnerabilities (detailed in \Cref {tab:target_apps}) to create eight API-sets(real API traffic datasets for individual APIs). Each API-set corresponds to a specific functional module explicitly linked to the CVEs listed in the \Cref {tab:target_apps}. Each API-set contains 500 traffic sequences (one sequence consists of multiple traffic records) that encompass normal operations, potentially risky behaviors, and concrete exploit attempts rooted in real-world BAC vulnerabilities. Given that constructing each traffic sequence in the dataset requires manual access and validation, we only built datasets for a total of 8 core APIs across these two projects. The dataset contains a total of 45,765 traffic records. With reference to the experimental scales of existing studies~\cite{li2024malicious, cui2023api2vec, galassi2005learning}, we contend that this scale is sufficient to enable robust methodology evaluation. Note that we did not model the global traffic of all APIs, but instead focused on functionally grouped API-sets, which can reduce noise from irrelevant traffic and improve the effectiveness of feature extraction~\cite{bovenzi2014level, hsiao2010cross}.

The basic information of these datasets is presented in~\Cref{tab:API-setOverview}. We categorize security-critical API traffic into two types: \textbf{violations}, which refers to all BAC violation traffic, including both intercepted and successfully escaped instances; and \textbf{exploits}, which denotes BAC violation traffic that has successfully escaped interception and effectively leverages known CVE vulnerabilities.

\subsection{Evaluation Metrics}

In BAC violation detection, the data distribution is often highly imbalanced, making evaluation metrics a critical part of performance assessment. 
We employ two categories of metrics to comprehensively evaluate our system: (1) \textbf{general metrics} to evaluate both API endpoint information identification and BAC violation detection, and (2) \textbf{specialized metrics} to address data imbalance and simulation quality.

We adopt \textbf{Accuracy (ACC)}, \textbf{Precision (P)}, \textbf{Recall (R)}, and the \textbf{F$_1$} as standard indicators for classification and detection tasks. 
Accuracy measures the overall proportion of correctly classified samples, Precision quantifies the correctness of predicted anomalies, and Recall measure the completeness of anomaly detection. 
The F$_1$ score, defined as the harmonic mean of Precision and Recall, balances the trade-off between detection completeness and reliability:
\begin{equation}
\text{F}_1 = 2 \times \frac{\text{P} \times \text{R}}{\text{P} + \text{R}}
\end{equation}

Considering that BAC violation detection inherently involves a severe class imbalance, we further employ the \textbf{Matthews Correlation Coefficient (MCC)} as a more stable and balanced indicator of classification quality. 
The MCC takes into account all four entries of the confusion matrix---true positives (TP), false positives (FP), true negatives (TN), and false negatives (FN)---and is computed as:
\begin{equation}
\mathrm{MCC} = 
\frac{
\mathrm{TP} \times \mathrm{TN} - \mathrm{FP} \times \mathrm{FN}
}{
\sqrt{
(\mathrm{TP} + \mathrm{FP})
(\mathrm{TP} + \mathrm{FN})
(\mathrm{TN} + \mathrm{FP})
(\mathrm{TN} + \mathrm{FN})
}
}
\end{equation}
Unlike ACC or F$_1$, the MCC provides a more comprehensive evaluation under class imbalance, which is crucial for assessing BAC violation detection models.

To further evaluate the quality of the LLM-based traffic simulation, we define the \textbf{API Coverage} metric ($\text{Cov}_{\text{API}}$) to jointly capture the diversity and uniformity of API usage:
\begin{equation}
\text{Cov}_{\text{API}} = \frac{|\mathcal{A}_{\text{used}}| / |\mathcal{A}|}{CV_{\text{API}}}
\end{equation}
where $|\mathcal{A}_{\text{used}}|$ denotes the number of distinct APIs used, $|\mathcal{A}|$ is the total number of APIs, and $CV_{\text{API}}$ is the coefficient of variation of API invocation frequency:
\begin{equation}
CV_{\text{API}} = \frac{\sqrt{\frac{1}{|\mathcal{A}|} \sum_{a \in \mathcal{A}} \left(f_a - \frac{1}{|\mathcal{A}|} \sum_{a \in \mathcal{A}} f_a\right)^2}}{\frac{1}{|\mathcal{A}|} \sum_{a \in \mathcal{A}} f_a}
\end{equation}
A higher $\text{Cov}_{\text{API}}$ indicates that simulated traffic covers a broader and more balanced range of API endpoints, contributing to more representative and robust training data. 
Additionally, we record the \textbf{execution time (Hour)} and \textbf{monetary cost (USD)} to assess overall system efficiency.

\subsection{Implementation Details}

All experiments were conducted on a workstation equipped with an Nvidia RTX 4090 graphics processing unit with 24 GB of memory and an Intel Xeon Platinum 8474C central processing unit with 16 GB of system memory. The data-generation component employed the \textbf{DeepSeek-R1} large language model agent. For sequence modeling, we used a Transformer encoder with an embedding dimension of 128, four attention heads, two encoder layers, and a feedforward size of 512. The model was trained for ten passes using the AdamW optimization algorithm with a learning rate of 1e-5. For the final classification stage, we trained the Gated CatBoost Network, which integrates a CatBoost decision-tree model (300 boosting iterations, depth of six, and a learning rate of 0.5) with a lightweight fully connected neural network whose hidden dimension is determined by the input size. The neural component was trained for eighty passes, and the outputs of the two experts were combined through a gating network with a small fully connected architecture that learns sample-specific fusion weights. These settings reflect the key numerical choices used in our implementation.

\section{Result Analysis}

\subsection{Comparison with SOTA BAC Detectors (RQ1)}

To evaluate the overall performance of \BAC, we compare it with four representative detection methods. IVD-HTTP is an invariant-based approach that detects access control violations by checking predefined request invariants. The other three baselines, LogAnomaly, BRM and BACAD, represent learning based detection methods that model user behavior or request patterns for anomaly identification. A brief description of each method is provided below.

\begin{table}[htbp]
\centering
\scriptsize
\vspace{-2.0ex}
\caption{Comparison of \BAC and Other Detection Methods Across Violation and Exploits Test Sets}
\label{tab:bac-vs-Detbaselines}
\setlength{\tabcolsep}{0.45mm}
\renewcommand\arraystretch{1.15}
\begin{tabular}{c|p{1.3cm}|ccccc|ccccc}
    \hline
    \multirow{2}{*}{\textbf{API-set}} & \multirow{2}{*}{\makecell{\textbf{Detection}\\\textbf{Methods}}}  
    & \multicolumn{5}{c|}{\textbf{Violation}} 
    & \multicolumn{5}{c}{\textbf{Exploits}} \\
    & & ACC & P & R & $\text{F}_1$ & MCC  
      & ACC & P & R & $\text{F}_1$ & MCC  \\
    \hline
    \multirow{6}{*}{API-set 1}

    & IVD-HTTP & 75.2 & 0.0 & 0.0 & 0.0 & -13.9 & 34.3 & 0.0 & 0.0 & 0.0 & -48.6 \\
    & LogAnomaly     & 87.5 & 32.7 & 35.6 & 34.0 & 27.1 & 86.2 &\cellcolor{gray!30}100.0 & 40.0 & 57.1 & 58.2 \\
    & BRM     & 83.2 & 16.0 & 16.0 & 16.0 & 6.7 & 22.9 & 16.0 & 40.0 & 22.9 & -44.0 \\
    & BACAD$_R$     & 85.0 & 36.6 & \cellcolor{gray!30}68.0 & 47.6 & 42.3 & 77.4 &\cellcolor{gray!30}100.0 & 40.0 & 57.1 & 56.8 \\
    & BACAD$_S$     & 84.0 & 33.3 & 60.0 & 42.9 & 36.4 & 73.3 &\cellcolor{gray!30}100.0 & 35.0 & 51.9 & 52.7 \\
& \BAC    & \cellcolor{gray!30}96.4 & \cellcolor{gray!30}100.0 & 64.0 & \cellcolor{gray!30}78.1 & \cellcolor{gray!30}78.5 & \cellcolor{gray!30}82.0 & \cellcolor{gray!30}100.0 & \cellcolor{gray!30}45.0 & \cellcolor{gray!30}62.1 & \cellcolor{gray!30}60.7 \\

    \hline
    \multirow{6}{*}{API-set 2}

    & IVD-HTTP & 77.8 & 27.4 & 74.0 & 40.0 & 35.3 & 32.9 & 28.6 & 90.0 & 43.4 & 0.0 \\
    & LogAnomaly     & 90.9 & 55.1 & 54.0 & 54.6 & 49.5 & 74.5 &\cellcolor{gray!30}100.0 & 40.0 & 57.1 & 58.2 \\
    & BRM     & 87.6 & 38.0 & 38.0 & 38.0 & 31.1 & 54.3 & 38.0 & 95.0 & 54.3 & 33.0  \\
    & BACAD$_R$     & 95.2 &\cellcolor{gray!30}100.0 & 52.0 & 68.4 & 70.3 & 76.0 &\cellcolor{gray!30}100.0 & 25.0 & 40.0 & 43.9 \\
    & BACAD$_S$     & 93.0 &\cellcolor{gray!30}100.0 & 30.0 & 46.2 & 52.8 & 65.0 & 0.0 & 0.0 & 0.0 & 0.0 \\

& \BAC    & \cellcolor{gray!30}98.6 & \cellcolor{gray!30}100.0 & \cellcolor{gray!30}86.0 & \cellcolor{gray!30}92.5 & \cellcolor{gray!30}92.0 & \cellcolor{gray!30}93.0 & \cellcolor{gray!30}100.0 & \cellcolor{gray!30}80.0 & \cellcolor{gray!30}88.9 & \cellcolor{gray!30}86.1 \\

    \hline
    \multirow{6}{*}{API-set 3}

    & IVD-HTTP & 80.6 & 10.2 & 12.0 & 11.0 & 0.2 & 68.6 & 40.0 & 20.0 & 26.3 & 10.3 \\
    & LogAnomaly     & 93.3 & 67.4 & 66.0 & 66.7 & 62.9 & 81.2 &\cellcolor{gray!30}100.0 & 30.0 & 46.2 & 47.8 \\
    & BRM     & 87.6 & 38.0 & 38.0 & 38.0 & 31.1 & 54.3 & 38.0 & \cellcolor{gray!30}95.0 & 54.3 & 33.0  \\
    & BACAD$_R$     & 97.4 & 97.4 & 76.0 & 85.4 & 84.8 & 87.9 &\cellcolor{gray!30}100.0 & 75.0 & 85.7 & 82.6 \\
    & BACAD$_S$     & 93.6 & 95.0 & 38.0 & 54.3 & 57.8 & 68.9 &\cellcolor{gray!30}100.0 & 30.0 & 46.2 & 48.4 \\

& \BAC    & \cellcolor{gray!30}99.2 & \cellcolor{gray!30}100.0 & \cellcolor{gray!30}92.0 & \cellcolor{gray!30}95.8 & \cellcolor{gray!30}95.5 & \cellcolor{gray!30}96.0 &\cellcolor{gray!30}100.0 & \cellcolor{gray!30}95.0 & \cellcolor{gray!30}97.4 & \cellcolor{gray!30}96.5 \\

    \hline
    \multirow{6}{*}{API-set 4}

    & IVD-HTTP & 72.0 & 11.9 & 28.0 & 16.7 & 3.5 & 67.1 & 41.2 & 35.0 & 37.8 & 15.8 \\
    & LogAnomaly     & \cellcolor{gray!30}95.4 & 77.6 & \cellcolor{gray!30}76.0 & \cellcolor{gray!30}76.8 & \cellcolor{gray!30}74.2 & \cellcolor{gray!30}86.8 & 83.3 & 25.0 & 38.5 & 36.3 \\
    & BRM     & 87.6 & 38.0 & 38.0 & 38.0 & 31.1 & 54.3 & 38.0 & \cellcolor{gray!30}95.0 & 54.3 & 33.0  \\
    & BACAD$_R$     & 93.2 &\cellcolor{gray!30}100.0 & 32.0 & 48.5 & 54.6 & 66.0 &\cellcolor{gray!30}100.0 & 35.0 & 51.9 & 52.7 \\
    & BACAD$_S$     & 91.4 &\cellcolor{gray!30}100.0 & 14.0 & 24.6 & 35.8 & 57.0 &\cellcolor{gray!30}100.0 & 5.0 & 9.5 & 19.0 \\

& \BAC    & 94.2 &\cellcolor{gray!30}100.0 & 42.0 & 59.2 & 62.8 & 71.0 &\cellcolor{gray!30}100.0 & 60.0 & \cellcolor{gray!30}75.0 & \cellcolor{gray!30}71.9 \\

    \hline
    \multirow{6}{*}{API-set 5}

    & IVD-HTTP & 88.6 & 46.2 & \cellcolor{gray!30}86.0 & 60.1 & 57.7 & 88.6 & 75.0 & 90.0 & 81.8 & 74.2 \\
    & LogAnomaly     & 91.3 & 57.1 & 56.0 & 56.6 & 51.7 & 75.6 & 66.7 & 20.0 & 30.8 & 24.8 \\
    & BRM     & 87.6 & 38.0 & 38.0 & 38.0 & 31.1 & 54.3 & 38.0 & \cellcolor{gray!30}95.0 & 54.3 & 33.0  \\
    & BACAD$_R$     & 93.2 & 94.4 & 34.0 & 50.0 & 54.4 & 66.9 &\cellcolor{gray!30}100.0 & 25.0 & 40.0 & 43.9 \\
    & BACAD$_S$     & 91.6 & 83.3 & 20.0 & 32.3 & 38.3 & 59.8 &\cellcolor{gray!30}100.0 & 5.0 & 9.5 & 19.0 \\

& \BAC    & \cellcolor{gray!30}98.6 &\cellcolor{gray!30}100.0 & \cellcolor{gray!30}86.0 & \cellcolor{gray!30}92.5 & \cellcolor{gray!30}92.0 & \cellcolor{gray!30}93.0 &\cellcolor{gray!30}100.0 & 90.0 & \cellcolor{gray!30}94.7 & \cellcolor{gray!30}93.0 \\

    \hline
     \multirow{6}{*}{API-set 6}

    & IVD-HTTP & 18.2 & 10.9 & \cellcolor{gray!30}100.0 & 19.6 & 10.0 & \cellcolor{gray!30}87.1 & 69.0 & \cellcolor{gray!30}100.0 & \cellcolor{gray!30}81.6 & 75.2 \\
    & LogAnomaly     & 93.7 & 69.4 & 68.0 & 68.7 & 65.2 & 82.3 & 83.3 & 33.3 & 47.6 & 45.6 \\
    & BRM     & 87.6 & 38.0 & 38.0 & 38.0 & 31.1 & 54.3 & 38.0 & 95.0 & 54.3 & 33.0  \\
    & BACAD$_R$     & 19.0 & 9.6 & 84.0 & 17.2 & -3.9 & 47.9 &\cellcolor{gray!30}100.0 & 65.0 & 78.8 & \cellcolor{gray!30}75.5 \\
    & BACAD$_S$     & 33.8 & 7.6 & 50.0 & 13.1 & -11.4 & 41.0 & 0.0 & 0.0 & 0.0 & 0.0 \\

& \BAC    & \cellcolor{gray!30}95.4 & \cellcolor{gray!30}84.6 & 66.0 & \cellcolor{gray!30}74.2 & \cellcolor{gray!30}72.3 & 82.3 &\cellcolor{gray!30}100.0 & 55.0 & 71.0 & 68.3 \\

    \hline
    \multirow{6}{*}{API-set 7}

    & IVD-HTTP & 28.6 & 28.6 & \cellcolor{gray!30}100.0 & 44.4 & 0.0 & 28.6 & 28.6 & \cellcolor{gray!30}100.0 & 44.4 & 0.0 \\
    & LogAnomaly     & 81.0 &\cellcolor{gray!30}100.0 & 33.9 & 50.6 & 51.7 & 66.9 &\cellcolor{gray!30}100.0 & 40.0 & 57.1 & 58.2 \\  
    & BRM     & 87.6 & 38.0 & 38.0 & 38.0 & 31.1 & 54.3 & 38.0 & 95.0 & 54.3 & 33.0  \\
    & BACAD$_R$     & 29.1 & 28.7 &\cellcolor{gray!30}100.0 & 44.6 & 4.4 & 50.3 & 28.9 &\cellcolor{gray!30}100.0 & 44.9 & 7.6 \\
    & BACAD$_S$     & 32.9 & 30.0 &\cellcolor{gray!30}100.0 & 46.0 & 13.4 & 53.0 & 30.3 &\cellcolor{gray!30}100.0 & 46.5 & 15.6 \\

& \BAC    & 89.0 & 72.8 & 98.3 & \cellcolor{gray!30}83.7 & \cellcolor{gray!30}77.6 & \cellcolor{gray!30}91.8 & 82.6 & 95.0 & \cellcolor{gray!30}88.4 & \cellcolor{gray!30}83.7 \\
    \hline
    \multirow{6}{*}{API-set 8}

    & IVD-HTTP & 10.0 & 10.0 & \cellcolor{gray!30}100.0 & 18.2 & 0.0 & 28.6 & 28.5 & \cellcolor{gray!30}100.0 & 44.4 & 0.0 \\
    & LogAnomaly     & 95.2 & 75.5 & 75.5 & 75.5 & 72.8 & 86.4 &\cellcolor{gray!30}100.0 & 30.0 & 46.2 & 47.8 \\
    & BRM     & 87.6 & 38.0 & 38.0 & 38.0 & 31.1 & 54.3 & 38.0 & 95.0 & 54.3 & 33.0  \\
    & BACAD$_R$     & 99.2 &\cellcolor{gray!30}100.0 & 92.0 & 95.8 & 95.5 & 96.0 &\cellcolor{gray!30}100.0 & 95.0 & 97.4 & 96.5 \\
    & BACAD$_S$     & \cellcolor{gray!30}99.2 &\cellcolor{gray!30}100.0 & 92.0 & \cellcolor{gray!30}95.8 & \cellcolor{gray!30}95.5 & \cellcolor{gray!30}96.0 &\cellcolor{gray!30}100.0 & \cellcolor{gray!30}95.0 & \cellcolor{gray!30}97.4 & \cellcolor{gray!30}96.5 \\

 & \BAC    & 96.2 &\cellcolor{gray!30}100.0 & 62.0 & 76.5 & 77.1 & 81.0 &\cellcolor{gray!30}100.0 & 75.0 & 85.7 & 82.6 \\

    \hline
\multirow{6}{*}{Average}
& IVD-HTTP & 56.4 & 18.2 & 62.5 & 26.3 & 11.6 & 54.5 & 38.9 & 66.9 & 45.0 & 15.9 \\
& LogAnomaly & 91.0 & 66.9 & 58.1 & 60.4 & 56.9 & 80.0 & 91.7 & 32.3 & 47.6 & 47.1 \\
& BRM & 87.1 & 35.3 & 35.3 & 35.3 & 28.1 & 50.4 & 35.3 & \cellcolor{gray!30}88.1 & 50.4 & 23.4 \\
& BACAD$_R$ & 76.4 & 70.8 & 67.3 & 57.2 & 50.3 & 71.1 & 91.1 & 57.5 & 62.0 & 57.4 \\
& BACAD$_S$ & 77.4 & 68.7 & 50.5 & 44.4 & 39.8 & 64.3 & 66.3 & 33.8 & 32.6 & 31.4 \\
&\BAC & \cellcolor{gray!30}96.0 & \cellcolor{gray!30}94.7 & \cellcolor{gray!30}74.5 & \cellcolor{gray!30}81.6 & \cellcolor{gray!30}81.0 & \cellcolor{gray!30}86.3 & \cellcolor{gray!30}97.8 & 74.4 & \cellcolor{gray!30}82.9 & \cellcolor{gray!30}80.4 \\
\hline
\end{tabular}
\vspace{-2.0ex}
\end{table}

\begin{itemize}

\item \textbf{IVD-HTTP:}
An invariant-based method that infers invariants of user-resource from historical HTTP logs. It works by extracting consistent access invariants and detecting violations when requests deviate from these learned patterns.

\item \textbf{LogAnomaly:}
A learning-based monitoring method that models API traffic as semantic sequences. It works by learning normal data-flow patterns and identifying anomalies when sequence consistency is disrupted.

\item \textbf{BRM:}
A learning-based BAC detection method that represents API traffic in a differential geometric space. It works by learning manifold structures of normal behaviors and detecting violations as geometric deviations.

\item \textbf{BACAD:}
A learning-based BAC detection method that extracts behavioral features from API traffic. It works by measuring geometric deviations from learned normal patterns to identify abnormal access behaviors.

\end{itemize}


\textbf{Regarding violation detection}, \BAC achieves an average absolute improvement of 21.2\% in $\text{F}_1$ and 24.1\% in MCC compared with the second-ranked method (one of the four baselines) for each API-set. 
BRM, which relies on per-request learning, lacks visibility into cross-request patterns; IVD-HTTP represents the invariant-based methods. It derives user–resource invariants from historical logs, so it cannot adapt to dynamic access semantics and often produces false alarms when real behaviors deviate from the inferred static patterns. Additionally, the other three baselines are learning based. LogAnomaly applies sequence modeling to identify abnormal data flows, yet it lacks semantic awareness across composite API interactions. BRM focuses on per-request representations and therefore misses cross-request behavioral dependencies. BACAD extracts statistical behavioral features but does not capture contextual relations across API calls. These limitations prevent learning based baselines from modeling composite API traffic effectively, while \BAC is explicitly designed to learn such patterns.

\textbf{Regarding exploit detection}, \BAC achieves an average absolute improvement of 20.9\% in $\text{F}_{1}$ and 23.0\% in MCC compared with the second-ranked method for each API set. Although BRM attains a relatively high average recall of 88.1\%, its low precision undermines user trust in BAC violation alerts and increases the operational burden on applications. Earlier SOTA methods show limited ability to model interdependent request sequences in composite API traffic, which are essential for detecting context-dependent access misuse. As a result, both baselines fail to provide practical exploit detection capability.

\begin{tcolorbox}[width=\linewidth, boxrule=1pt, sharp corners=all,
left=2pt, right=2pt, top=2pt, bottom=2pt, colback=white, colframe=black]
\textbf{Answer to RQ1}: \BAC shows the strongest overall performance among all SOTA baselines.
For violation detection, it achieves superior results (F1: 81.6\%, MCC: 81.0\%).
For exploit detection, it again leads (F1: 82.9\%, MCC=80.4) while maintaining high precision,
enabling integration with existing access control defenses without impacting benign users.
\end{tcolorbox}

\subsection{Effectiveness of API Information Extraction (RQ2)}

To evaluate the effectiveness of \textbf{Phase~1 of \BAC} in discovering API endpoint information from raw traffic, we collected an annotated dataset during the acceptance phase of a one-year university–industry collaboration.

\begin{table}[htbp]
    \centering
    \scriptsize
    \setlength{\tabcolsep}{0.8mm} 
    \renewcommand\arraystretch{1.2}
    \caption{Expert Information}
    \label{tab:expert_info}
    \begin{threeparttable}
        \begin{tabular}{l|p{2cm}|l|p{3cm}}
            \hline
            \textbf{Expert} & \textbf{Title} & \textbf{Experience} & \textbf{Field} \\
            \hline
            Expert 1 & Senior Engineer & 5 years & Data and PaaS security \\
            Expert 2 & Senior Engineer & 23 years & Data and PaaS security \\
            Expert 3 & Assistant Professor & 7 years & Security of software and AI \\
            \hline
        \end{tabular}
    \end{threeparttable}
\end{table}

We invited the domain experts listed in \Cref{tab:expert_info} to guide two master’s students and two software engineers, who independently annotated the extracted API information and then cross-checked and reconciled their annotations as part of the acceptance process. The process involved approximately 50 hours of expert labor. For comparison, we selected several representative unsupervised clustering methods applied to unlabeled network traffic, and the brief descriptions of each method is provided below.

\begin{itemize}
    \item \textbf{K-means:} A partition-based clustering method that groups requests by vectorized feature similarity (e.g., tokenized paths and parameters). Its reliance on statistical similarity makes it a simple baseline for separating API endpoints with distinct structural patterns.
    
    \item \textbf{DBSCAN:} A density-based clustering method that groups closely packed requests while treating isolated ones as noise. Its ability to detect clusters of arbitrary shape helps capture irregular or low-frequency API access patterns that deviate from the main traffic.
    
    \item \textbf{Spectral:} A graph-based method that builds a similarity graph over requests and partitions it using Laplacian eigenvectors. By leveraging pairwise structural relations, it can reveal request groups that share consistent API semantics.
\end{itemize}

\begin{figure}[htbp]
    \centering
    \includegraphics[width=1.0\linewidth]{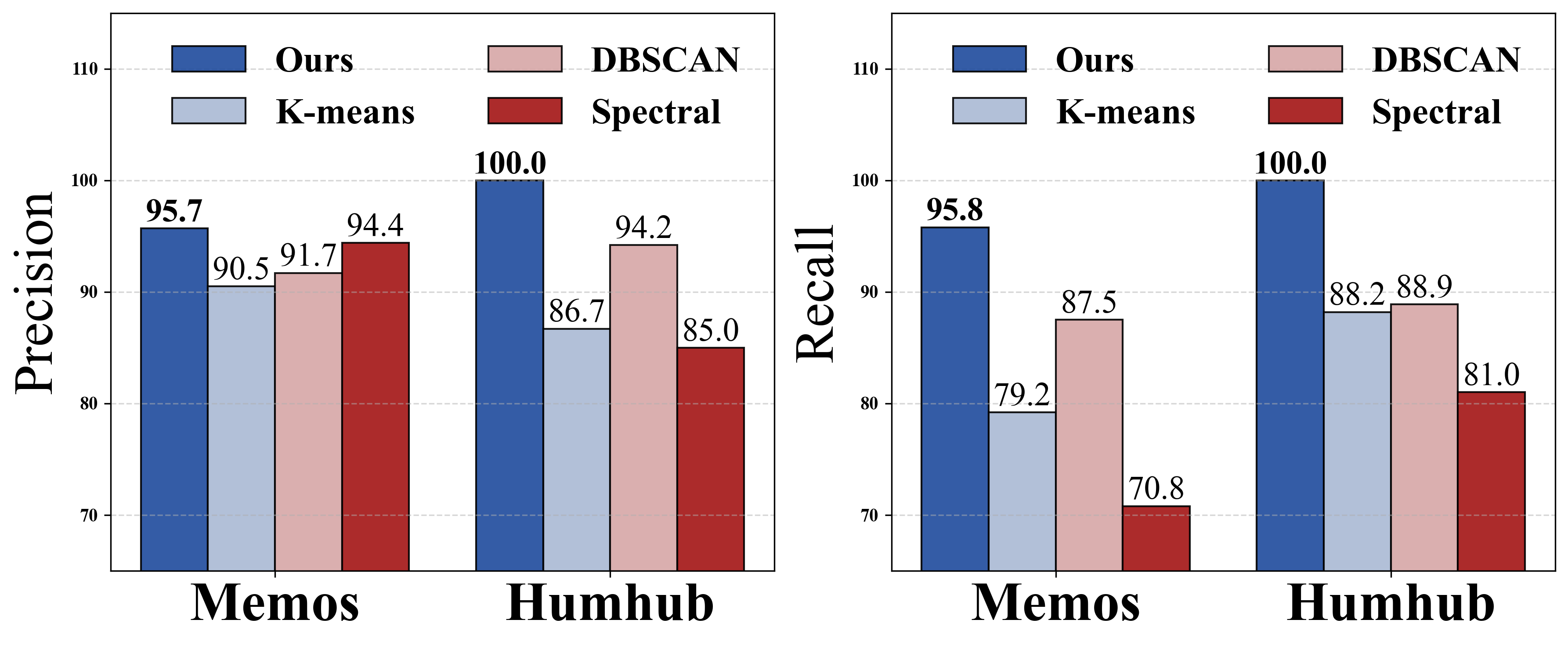}
        \caption{Performance Comparison on API Endpoint Discovery Using Precision and Recall Metrics.}
    \label{fig: API_Discovery}
    \vspace{-1.0ex}
\end{figure}

For a complete list of API endpoints, we evaluate each method using precision and recall. As shown in~\figurename~\ref{fig: API_Discovery}, Phase~1 of \BAC consistently outperforms all baselines on both metrics.

\textbf{On precision}, \BAC achieves 95.7\% on \textbf{Memos} and 100\% on \textbf{Humhub}, exceeding the second-best baselines by 1.3\% and 5.8\%, respectively. This higher precision indicates that \BAC can more accurately distinguish true API endpoints from structurally similar but semantically unrelated traffic.

\textbf{On recall}, \BAC reaches 95.8\% on \textbf{Memos} and 100\% on \textbf{Humhub}, outperforming the next-best baselines by 8.3\% and 11.1\%. The higher recall demonstrates that \BAC can recover a more complete set of API endpoints, including infrequent or sparsely observed ones that clustering baselines tend to miss.

\begin{tcolorbox}[width=\linewidth, boxrule=1pt, sharp corners=all,
  left=2pt, right=2pt, top=2pt, bottom=2pt, colback=white, colframe=black]
  \textbf{Answer to RQ2}: Phase~1 of \BAC achieves the best overall performance, with both precision and recall consistently exceeding 95\% across datasets. These results show that our approach can reliably extract complete and accurate API endpoint information from unlabeled traffic, indicating that it can replace manual efforts.
\end{tcolorbox}

\subsection{Effectiveness of Simulated Data Generation (RQ3)}


To evaluate the effectiveness of Phase~2 of \BAC in simulating realistic API traffic, we select six popular LLMs as candidate API Traffic Generators and 2 simulation methods used in API traffic generation. The compared methods are introduced as follows.

\begin{itemize}

\item \textbf{Qwen-Max:} 
A 110B mixture-of-experts language model released by Alibaba Cloud. It dynamically activates expert subnetworks for each task, enabling efficient computation and strong performance across diverse NLP benchmarks.

\item \textbf{Llama3-405B:} 
A 405B-parameter model from Meta optimized for large-scale deployment. It offers improved contextual comprehension, long-form reasoning, and multilingual understanding.

\item \textbf{DeepSeek-R1:} 
A 671B-parameter mixture-of-experts model from DeepSeek AI, featuring strong long-context modeling, robust reasoning, and efficient expert routing.

\item \textbf{GPT-4o:} 
A 200B-class OpenAI model optimized for faster inference while maintaining strong reasoning, comprehension, and multimodal performance.

\item \textbf{o1-preview:} 
An OpenAI model enhanced via reinforcement learning to approximate human-like reasoning trajectories, improving coherence and problem-solving abilities.

\item \textbf{GPT-5:} 
The next-generation flagship model from OpenAI, designed with advanced reasoning, stronger long-context understanding, and improved performance across multimodal and agentic tasks.

\item \textbf{IGDF:} 
An iterative data generation framework combining large language models with dual-channel filtering. It alternates between LLM-based sample generation and expert/pseudo-model filtering to produce high-quality synthetic API request datasets.

\item \textbf{VSF:} 
A simulation-based data generation method that constructs synthetic API behaviors by modeling variable states and function flows. It creates controllable and diverse request patterns for evaluating anomaly detection under structured scenarios.

\end{itemize}

\begin{table}[htbp]
\scriptsize
\vspace{-2.0ex}
\centering
\caption{Evaluation of API Coverage and Generation Efficiency Across Different Data Generation Methods}
\label{tab:llm_API-set_evaluation_reduced}
\setlength{\tabcolsep}{1.5mm} 
\renewcommand\arraystretch{1.2}  
\begin{threeparttable}
\begin{tabular}{c|ccc|ccc}
    \hline
    \multirow{2}{*}{LLM} 
        & \multicolumn{3}{c|}{\textbf{Humhub}} 
        & \multicolumn{3}{c}{\textbf{Memos}} \\
    & $\text{Cov}_{\text{API}} $ & $USD$ & $Hour$ & $\text{Cov}_{\text{API}} $ & $USD$ & $Hour$ \\
    \hline
    Qwen-Max      & 139.0 & 109.8 & 0.7 & 158.3 & 62.4 & 0.8 \\
    Llama3        & 131.7 & 6.6   & 0.8 & 134.0 & 3.8  & 0.7 \\
    DeepSeek-R1   & 141.7 & 4.1   & 10.1 & 214.2 & 2.8 & 8.9 \\
    GPT-4o        & 140.8 & 98.8  & 0.5 & 176.2 & 56.3 & 0.4 \\
    O1-preview    & 153.5 & 394.7 & 0.4 & 207.7 & 225.3 & 0.4 \\
    GPT-5         & 143.6 & 99.3  & 0.5 & 181.3 & 61.3 & 0.4 \\
    \hline
    IGDF          & 91.7  & 23.1  & 20.1 & 96.0 & 25.3 & 15.2 \\
    VSF$^\dagger$           & 131.7 & -     & 0.4  & 134.0 & -   & 0.5 \\
    Ours          & \cellcolor{gray!30}162.5 & \cellcolor{gray!30}3.7 & 10.9  
                  & \cellcolor{gray!30}249.7 & \cellcolor{gray!30}2.1 & 9.2 \\
    \hline
\end{tabular}
\begin{tablenotes}
\footnotesize
\item[\textdagger] VSF performs variable substitution without invoking any LLM, so it incurs no additional cost.
\end{tablenotes}
\end{threeparttable}
\end{table}

\begin{table*}[htbp]
\scriptsize
\centering
\caption{Performance of ADMs Trained with Different Simulation Methods on Humhub and Memos}

\label{tab:simulation-performance}
\setlength{\tabcolsep}{1.0pt} 
\renewcommand\arraystretch{1.5} 
\begin{threeparttable}
\begin{tabular}{
c|
ccccc|ccccc|ccccc|
ccccc|ccccc|ccccc}
\hline
\multirow{3}{*}{\textbf{{\makecell{\textbf{Data}\\\textbf{Generator}}} }} &
\multicolumn{15}{c|}{\textbf{Humhub}} &
\multicolumn{15}{c}{\textbf{Memos}}\\
\cline{2-31}
& \multicolumn{5}{c|}{Anomaly Detection} & \multicolumn{5}{c|}{Statistical Ensemble} & \multicolumn{5}{c|}{Heterogeneous Ensemble}
& \multicolumn{5}{c|}{Anomaly Detection} & \multicolumn{5}{c|}{Statistical Ensemble} & \multicolumn{5}{c}{Heterogeneous Ensemble}\\
\cline{2-6}\cline{7-11}\cline{12-16}\cline{17-21}\cline{22-26}\cline{27-31}
& ACC & P & R & F1 & MCC
& ACC & P & R & F1 & MCC
& ACC & P & R & F1 & MCC
& ACC & P & R & F1 & MCC
& ACC & P & R & F1 & MCC
& ACC & P & R & F1 & MCC\\
\hline

Qwen-Max & 80.8 & 27.0 & 54.0 & 36.0 & 28.3 & \cellcolor{gray!20}96.1 & \cellcolor{gray!20}80.9 & \cellcolor{gray!20}80.0 & \cellcolor{gray!20}80.5 & \cellcolor{gray!20}78.3 & 95.0 & 100.0 & 49.6 & 66.3 & 68.5
& \cellcolor{gray!20}84.6 & \cellcolor{gray!20}44.6 & \cellcolor{gray!20}67.5 & \cellcolor{gray!20}53.7 & \cellcolor{gray!20}46.4 & 73.1 & 27.9 & 65.0 & 39.0 & 28.9 & 76.9 & 31.1 & 61.9 & 41.4 & 31.6\\

Llama3 & 81.8 & 29.4 & 58.8 & 39.2 & 32.3 & \cellcolor{gray!20}95.4 & \cellcolor{gray!20}80.7 & \cellcolor{gray!20}70.4 & \cellcolor{gray!20}75.2 & \cellcolor{gray!20}72.9 & 94.2 & 87.9 & 49.2 & 63.1 & 63.2
& \cellcolor{gray!20}80.5 & \cellcolor{gray!20}34.3 & \cellcolor{gray!20}51.9 & \cellcolor{gray!20}41.3 & \cellcolor{gray!20}31.1 & 49.5 & 10.4 & 36.9 & 16.2 & -7.9 & 69.3 & 14.7 & 27.5 & 19.1 & 2.5\\

DeepSeek-R1 & 75.9 & 14.8 & 29.6 & 19.7 & 8.0 & 87.4 & 43.7 & 91.2 & 59.1 & 57.7 & \cellcolor{gray!20}88.7 & \cellcolor{gray!20}46.6 &  \cellcolor{gray!20}\underline{\textbf{91.2}} & \cellcolor{gray!20}61.7 &\cellcolor{gray!20}60.2
& 72.1 & 13.2 & 20.0 & 15.9 & 0.0 & \cellcolor{gray!20}44.3 & \cellcolor{gray!20}18.4 & \cellcolor{gray!20}93.8 & \cellcolor{gray!20}30.8 & \cellcolor{gray!20}22.0 & 43.2 & 17.9 &  91.9 & 30.0 & 20.1
\\  
GPT-4o & 82.9 & 32.2 & 64.4 & 42.9 & 37.0 & 93.2 & 100.0 & 32.4 & 48.9 & 54.9 & \cellcolor{gray!20}95.0 & \cellcolor{gray!20}100.0 & \cellcolor{gray!20}49.6 & \cellcolor{gray!20}66.3 & \cellcolor{gray!20}68.5
& 87.4 & 51.7 & 78.1 & \cellcolor{gray!20}62.2 & \cellcolor{gray!20}56.7 & 62.3 & 22.1 & 73.1 & 33.9 & 23.1 & 78.4 & 32.9 & 61.3 & 42.8 & 33.2\\
o1-preview & 80.3 & 25.8 & 51.6 & 34.4 & 26.3 & \cellcolor{gray!20}96.2 & \cellcolor{gray!20}100.0 & \cellcolor{gray!20}62.4 & \cellcolor{gray!20}76.9 & \cellcolor{gray!20}77.4 & 96.1 & 100.0 & 61.2 & 75.9 & 76.6
& \cellcolor{gray!20}88.9 & \cellcolor{gray!20}55.4 & \cellcolor{gray!20}83.8 & \cellcolor{gray!20}66.7 & \cellcolor{gray!20}62.2 & 88.5 & 100.0 & 13.1 & 23.2 & 34.1 & 90.1 & 62.5 & 62.5 & 62.5 & 56.8
\\
GPT-5 & 75.6 & 14.0 & 28.0 & 18.7 & 6.7 & 89.6 & 48.9 & 81.2 & 61.1 & 57.9 & \cellcolor{gray!20}91.3 & \cellcolor{gray!20}54.5 & \cellcolor{gray!20}79.6 & \cellcolor{gray!20}64.7 & \cellcolor{gray!20}61.4
& 76.5 & 24.4 & 36.9 & 29.4 & 16.5 & 44.1 & 17.4 & 86.3 & 29.0 & 17.1 & \cellcolor{gray!20}42.6 & \cellcolor{gray!20}18.6 & \cellcolor{gray!20}\textbf{\underline{98.8}} & \cellcolor{gray!20}31.3 & \cellcolor{gray!20}24.3\\

\hline
\hline
IGDF$^\S$ & \cellcolor{gray!20}81.2 & \cellcolor{gray!20}28.0 & \cellcolor{gray!20}56.0 & \cellcolor{gray!20}37.3 & \cellcolor{gray!20}30.0 & - & - & - & - & - & - & - & - & - & - & \cellcolor{gray!20}89.9 & \cellcolor{gray!20}57.9 & \cellcolor{gray!20}87.5 & \cellcolor{gray!20}69.7 & \cellcolor{gray!20}65.9 & - & - & - & - & - & - & - & - & -& -
\\
VSF & \cellcolor{gray!20}75.5 & \cellcolor{gray!20}13.8 & \cellcolor{gray!20}27.6 &\cellcolor{gray!20}18.4 & \cellcolor{gray!20}6.3 & 21.2 & 1.2 & 8.8 & 2.2 & -45.1 & 23.3 & 1.6 & 10.8 & 2.7 & -41.8
& \cellcolor{gray!20}88.9 & \cellcolor{gray!20}55.4 & \cellcolor{gray!20}83.8 & \cellcolor{gray!20}66.7 & \cellcolor{gray!20}62.2 & 10.3 & 2.9 & 17.5 & 4.9 & -63.4 & 28.7 & 1.5 & 6.9 & 2.5 & -42.2
\\
Ours$^\ddag$ & 82.7 & 31.8 & 63.6 & 42.4 & 36.3 & 96.4 & 78.7 & 87.2 & 82.7 & 80.8 &  \cellcolor{gray!20}\underline{\textbf{97.2}} & \cellcolor{gray!20}\underline{\textbf{100.0}} & \cellcolor{gray!20}72.4 &  \cellcolor{gray!20}\underline{\textbf{84.0}} &  \cellcolor{gray!20}\underline{\textbf{83.9}}
& 89.3 & 56.2 & 85.0 & 67.7 & 63.4 & 55.5 & 21.8 & 91.9 & 35.3 & 28.5 &  \cellcolor{gray!20}\textbf{\underline{94.4}} &  \cellcolor{gray!20}\textbf{\underline{79.9}} & \cellcolor{gray!20}76.9 &  \cellcolor{gray!20}\textbf{\underline{78.3}} &  \cellcolor{gray!20}\textbf{\underline{75.1}}\\

\hline
\end{tabular}
\begin{tablenotes}
\footnotesize
\item[\S] Since IGDF lacks malicious annotations, it is limited to training anomaly detection models.
\item[\ddag] Our method uses DeepSeek-R1 as the backbone, with the \BAC-based retrieval-augmented and a hallucination-elimination phase integrated into the API traffic generation pipeline.
\end{tablenotes}
\end{threeparttable}
\vspace{-2.0ex}

\end{table*}

\textbf{In terms of generation efficiency}, DeepSeek-R1 achieves the highest $\text{Cov}_{\text{API}}$ in both applications, with scores of 162.5 for Humhub and 249.7 for Memos, as shown in~\Cref{tab:llm_API-set_evaluation_reduced}. This result demonstrates that DeepSeek-R1 generates the most diverse and balanced API traffic, likely attributed to its longer reasoning chain. However, this advantage is accompanied by significantly long data generation time consumption, exceeding 10 hours, compared with under 1 hour for other models. Despite the long-term cost, DeepSeek-R1 remains highly cost-efficient, with the costs of only \$3.7 for Humhub and \$2.1 for Memos -- approximately 4\% of the cost required by GPT-5.

\textbf{In terms of detection performance}, we evaluate how well different simulation methods support the training of anomaly detectors by testing three anomaly detection methods: Anomaly Detection (VAE-CNN~\cite{chen2018autoencoder}), Statistical Ensemble (Random Forest~\cite{lin2020anomaly}), and Heterogeneous Ensemble (GateCatboostNet, Cf.~\Cref{Met: detector}). 
As shown in \Cref{tab:simulation-performance}, the data generated by our method consistently leads to strong performance across both Humhub and Memos. 

On Humhub, our Heterogeneous Ensemble Model trained on simulated data achieves an MCC of 84.0\% and an $\text{F}_1$ of 83.9\%, surpassing the second-best baseline (the Statistical Ensemble Model trained on Qwen-Max samples) by 3.5 and 5.6 percentage points, respectively. Besides, our model also attains an accuracy of 97.2\%, a precision of 100.0\%, and a recall of 72.4\%, collectively reflecting a consistently strong and well-balanced detection capability. 

On Memos, our method achieves an MCC of 78.3\% and an $\text{F}_1$ of 75.1\%, surpassing the second-highest results (from the VAE-CNN model trained on IGDF samples) by 8.6\% in MCC and 9.2\% in $\text{F}_1$. Additionally, our approach attains an accuracy of 94.9\%, a precision of 79.9\%, and a recall of 76.9\%, demonstrating strong overall performance across all evaluation dimensions. These results show that our simulated data enables substantially more effective generalization, capturing a broader and more balanced range of behavioral patterns than other data generators.

\begin{tcolorbox}[width=\linewidth, boxrule=1pt, sharp corners=all, left=2pt, right=2pt, top=2pt, bottom=2pt, colback=white, colframe=black]
  \textbf{Answer to RQ3}: The BAC violation detector trained on the simulated data produced by our Phase~2 simulator achieves the best detection performance on both Humhub and Memos. While the simulator is slower than rule-based generation, its offline nature and low marginal cost make it a practical and reliable approach for producing high-quality traffic when accurate BAC modeling is required.
\end{tcolorbox}

   \vspace{-1.0ex}

\subsection{Effectiveness of Sequential Feature (RQ4)}

\begin{figure}[htbp]
    \centering

    \includegraphics[width=1.0\linewidth]{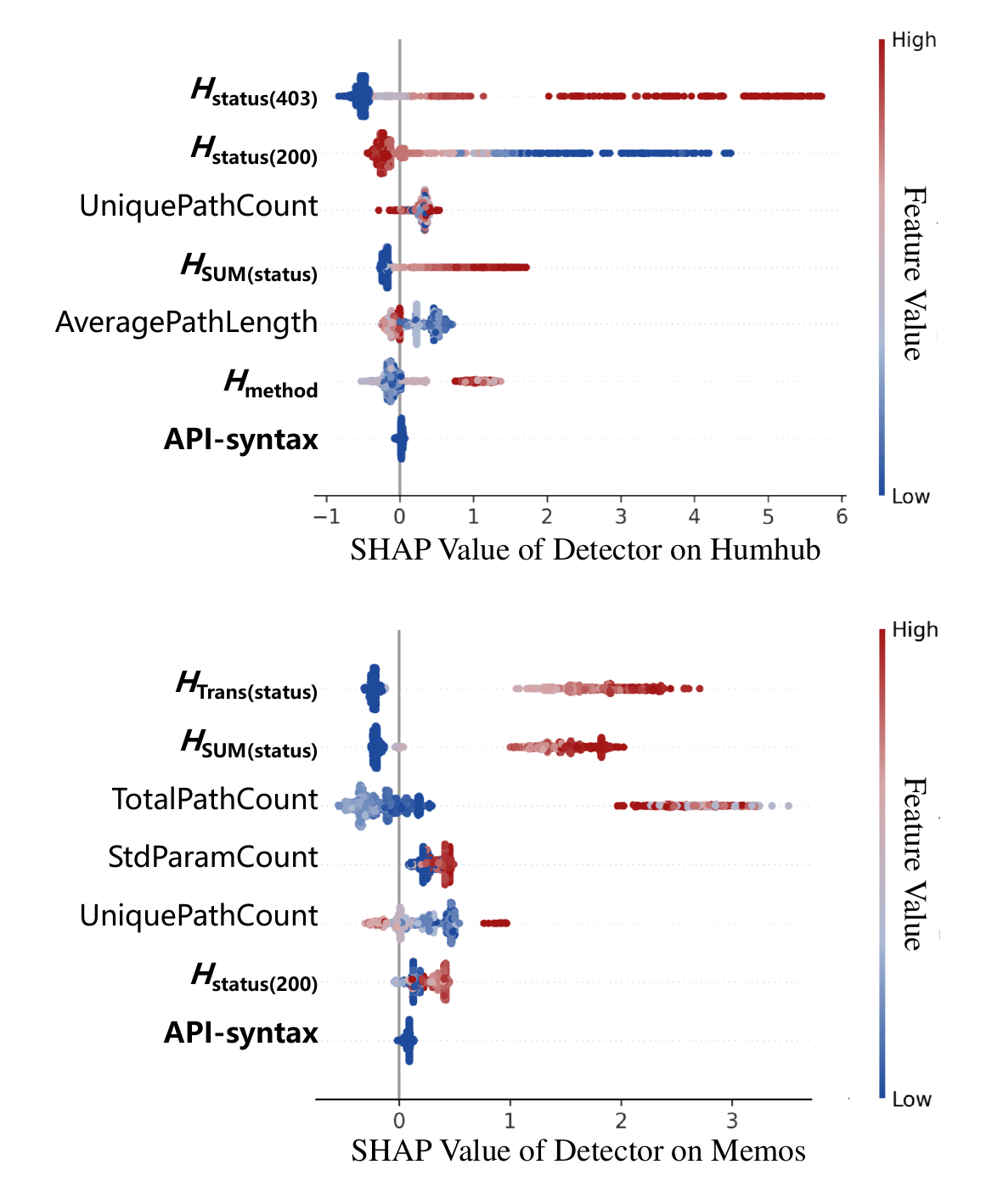}

    \caption{SHAP Summary Value of Feature Importance Across 2 Project.}

    \label{fig: shap-summary}
\end{figure}

To investigate whether the sequential features introduced in \BAC truly support correct classification, we examine their SHAP contributions on both Humhub and Memos, as shown in~\figurename~\ref{fig: shap-summary}. 

For Humhub, entropy-based features of sequential features introduced by us such as $H_{\text{status(403)}}$, $H_{\text{status(200)}}$, and $H_{\text{SUM(status)}}$ consistently rank among the most influential predictors, reflecting the strong discriminative power of status-code concentration and uncertainty patterns within its complex interaction flows. Additionally, for Memos, sequential features also play a central role, with $H_{\text{Trans(status)}}$ and $H_{\text{SUM(status)}}$ emerging as key contributors that capture the temporal variations and distributional shifts in status-code sequences. The prominence of these features in both systems demonstrates that the sequential characteristics we introduced are broadly effective: regardless of system complexity, modeling status-code dynamics across request sequences provides stable and substantial benefits for BAC detection, enabling the model to recognize context-dependent anomalies that are not detectable through single-request features alone.

\begin{tcolorbox}[width=\linewidth, boxrule=1pt, sharp corners=all,
  left=2pt, right=2pt, top=2pt, bottom=2pt, colback=white, colframe=black]
\textbf{Answer to RQ4}: The sequential features introduced in Phase~3 of \BAC are highly effective for detecting BAC violations. Across both Humhub and Memos, SHAP analysis reveals that our two newly designed sequential feature types consistently rank among the most influential predictors and play a critical role in detecting BAC violations.
\end{tcolorbox}

\subsection{Ablation Study (RQ5)}

To investigate the contribution of each component in \BAC, we conduct comprehensive ablation studies by separately removing the RAG (\textbf{AG}), the Hallucination Elimination (\textbf{HE}), and the Sequential Features (\textbf{SF}). The evaluation is carried out on two benchmark applications, \textbf{Humhub} and \textbf{Memos}. The detailed results are presented in Table~\ref{tab:ablation}.

\begin{table}[htbp]
\centering
\scriptsize
\vspace{-2.0ex}
\caption{The Effectiveness of Each Component in \BAC.}
\label{tab:ablation}
\setlength{\tabcolsep}{1mm}
\renewcommand\arraystretch{1.15}
\begin{tabular}{c|c|ccccc}
\hline
\textbf{App} & \textbf{Setting} & \textbf{ACC} & \textbf{P} & \textbf{R} & $\textbf{F}_1$ & \textbf{MCC} \\ 
\hline
\multirow{3}{*}{Humhub} 
& w/o-AG & 95.8($\downarrow$1.7) & 94.5($\downarrow$5.5) & 62.0($\downarrow$10.4) & 74.9($\downarrow$9.1) & 74.6($\downarrow$9.4) \\
& w/o-HE & 92.5($\downarrow$4.9) & 58.2($\downarrow$41.8) & 89.6($\uparrow$17.2) & 70.6($\downarrow$13.9) & 68.5($\downarrow$15.3) \\
& w/o-SF & 91.0($\downarrow$6.1) & 67.1($\downarrow$32.9) & 19.6($\downarrow$52.8) & 30.3($\downarrow$53.6) & 33.0($\downarrow$50.5) \\
\hline
\multirow{3}{*}{Memos} 
& w/o-AG & 45.0($\downarrow$49.4) & 19.2($\downarrow$60.6) & 98.8($\uparrow$22.0) & 32.2($\downarrow$46.1) & 25.8($\downarrow$49.6) \\
& w/o-HE & 47.4($\downarrow$47.0) & 18.8($\downarrow$61.0) & 90.0($\uparrow$13.1) & 31.2($\downarrow$47.2) & 21.7($\downarrow$53.9) \\
& w/o-SF & 91.7($\downarrow$2.7) & 67.9($\downarrow$12.0) & 70.0($\downarrow$6.9) & 68.9($\downarrow$9.5) & 64.1($\downarrow$11.7) \\
\hline
\end{tabular}

\begin{flushleft}
\footnotesize
\justify
“w/o-AG”, “w/o-HE”, and “w/o-SF” denote the removal of RAG, Hallucination Elimination, and Sequential Features, respectively.
\end{flushleft}
   \vspace{-1.0ex}
\end{table}

As shown in Table~\ref{tab:ablation}, removing any component leads to a performance decline, validating the necessity of each component. Notably, excluding the Hallucination Elimination (w/o-HE) results in the most significant decline on Humhub, with precision dropping by 41.8, F\textsubscript{1} by 13.9, and MCC by 15.3, highlighting its crucial role in ensuring data reliability. The performance degradation from removing the RAG (w/o-AG) is relatively minor, with F\textsubscript{1} decreasing by 9.1 and MCC by 9.4, possibly because DeepSeek-R1 supports long input contexts (up to 128k tokens)~\cite{guo2025deepseek}, which reduces dependence on external RAG techniques~\cite{r1-No-RAG}. The Sequential Features (SF) is also essential, as its removal (w/o-SF) causes substantial declines in precision (32.9), recall (52.8), F\textsubscript{1} (53.6), and MCC (50.5), demonstrating the importance of modeling temporal patterns in API traffic sequences.

On Memos, removing AG (w/o-AG) leads to large decreases in F\textsubscript{1} (46.1) and MCC (49.6), despite recall increasing due to overprediction. Similarly, removing HE (w/o-HE) reduces F\textsubscript{1} by 47.2 and MCC by 53.9, again confirming its role in improving data correctness. Removing SF (w/o-SF) also impacts performance, with declines of 9.5 in F\textsubscript{1} and 11.7 in MCC, showing that sequential information remains important even in this application.

\begin{tcolorbox}[width=\linewidth, boxrule=1pt, sharp corners=all,
left=2pt, right=2pt, top=2pt, bottom=2pt, colback=white, colframe=black]
\textbf{Answer to RQ5}: The ablation study shows that all core components of \BAC are indispensable, as removing AG, HE, or SF consistently results in notable performance degradation across both Humhub and Memos.
\end{tcolorbox}

\section{Discussion}
\label{sec: discussion}

\subsection{Complementarity with Existing Built-in Protection Mechanisms}

In practical enterprise settings, \BAC is deployed as an additional protection layer on top of existing access control mechanisms. Since our approach relies on collecting API traffic within a time window to extract behavioral features, it is typically applied only to security-critical APIs to limit overhead and avoid unnecessary false-positive handling. The framework also supports flexible intervention strategies. Organizations can choose how to respond to suspicious activity based on its frequency, rather than enforcing immediate blocking. In our university–industry collaborations, this often means starting with mild actions such as warnings or temporary rate limits, and escalating to stricter blocking only when suspicious behavior persists. This design allows \BAC to strengthen existing defenses while keeping operational cost and user impact low.

To evaluate the practical effectiveness of \BAC, we deployed it alongside the built-in access control mechanisms of two applications, Humhub and Memos. This configuration mirrors real-world scenarios where \BAC acts as a complementary layer rather than a replacement for existing security protections. The results are presented in Table~\ref{table: complementarity}.

\begin{table}[htbp]
\centering
\vspace{-2.0ex}
\scriptsize
\caption{Performance Comparison of Built-in Protection Mechanisms with and without \BAC.}
\label{table: complementarity}
\setlength{\tabcolsep}{1.2mm}
\renewcommand\arraystretch{1.15}
\begin{tabular}{c|c|cccccc}
\hline
Application & Detection & ACC & P & $R$ & $\text{F}_1$ & MCC & FPR \\ \hline
\multirow{2}{*}{Humhub}
& Built-in Only & 96.0 & 100.0 & 60.0 & 75.0 & 75.8 & 0.0\\
& Built-in + \BAC & \textbf{99.6} & \textbf{100.0} & \textbf{96.0} & \textbf{97.9} & \textbf{97.7} & 0.0\\ \hline

\multirow{2}{*}{Memos}
& Built-in Only & 96.0 & 100.0 & 60.0 & 75.0 & 75.8 & 0.0\\
& Built-in + \BAC & 89.1 & 72.8 & \textbf{98.3} & \textbf{83.7} & \textbf{77.7} & 14.7\\ \hline
\end{tabular}
\vspace{-1.0ex}
\end{table}

For Humhub, the standalone built-in mechanism achieves an ACC of 96.0\%, with a perfect Precision at 100.0\% but a low Recall at 60.0\%, resulting in an $\text{F}_1$ of 75.0\%.
This indicates significant undetected violations. After integrating \BAC, ACC increases to 99.6\%, Recall improves drastically to 96.0\%, and $\text{F}_1$ reaches 97.9\%. Notably, Precision remains at 100.0\%, and False Positive Rate (FPR) stays at 0.0\%. 

In the case of Memos, integrating \BAC drives Recall to 98.3\% and boosts \(\text{F}_1\) to 83.7\%. While Precision drops to 72.8\% and FPR rises to 14.7\%, MCC still increases from 75.8\% to 77.7\%, reflecting a more balanced and effective detection capability overall.

These results show that \BAC consistently enhances detection across different applications. It improves recall and overall effectiveness, especially in identifying BAC violations that built-in mechanisms tend to miss.


\subsection{Simulated Data Enables BAC Violation Detection}

Our research confirms the feasibility of generating simulated data for BAC issues based on LLMs and using it for the training of detection models. BAC violations typically arise from cross-role or cross-session sequences rather than individual requests, rendering them underrepresented in real data distributions. Leveraging domain-specific knowledge, LLMs can systematically construct these composite behaviors, enabling models to learn access transitions in low-density or unseen regions of the feature space. The simulation enhances generalization and facilitates decision boundary learning~\cite{qing2024low}. Additionally, simulated composite API traffic preserves the statistical properties and feature schema of real traffic, ensuring compatibility with existing detection pipelines. By injecting precise labels for each access context, simulation facilitates supervised and semi-supervised training on structural patterns that are otherwise ambiguous, enabling robust learning of temporal and relational indicators critical to BAC violation detection~\cite{jan2020throwing}.
\begin{table}[htbp]
\centering
\scriptsize
\vspace{-4.0ex}
\caption{Evaluation of BAC Detector Trained by Real API Traffic Performance Using 10-Fold Cross-Validation.}
\label{table:humhub_gap_updated}
\renewcommand\arraystretch{1.3}
\setlength{\tabcolsep}{1.8mm}
\begin{tabular}{c|cccccc}
\hline
Application & ACC & P & R & $\text{F}_1$ & MCC  \\ \hline

Humhub & 
97.6(1.2) & 
89.7(5.5) & 
93.5(5.5) & 
91.4(4.1) & 
90.1(4.8) &  \\

Memos & 97.1(1.8) & 90.8(9.6) & 90.3(7.8) & 90.0(5.6) & 88.7(6.2) \\
\hline
\end{tabular}
\vspace{-2.0ex}
\end{table}

To further explore the gap between training models with simulated data and real data, we trained an Ensemble model using the test data introduced in Section~\ref{sec:dataset} and conducted 10-fold cross-validation. Specifically, We treat the test data as real data because it is a dataset established through manual access based on CVE reports. In the 10-fold validation, we randomly divided the real data, with 90\% used as the training set and 10\% as the test set. The results are shown in~\Cref{table:humhub_gap_updated}. By comparing the performance of \BAC presented in~\Cref{tab:simulation-performance} with these results, we observe a discrepancy between using simulated data and directly using real data for model training, indicating that \BAC still has room for improvement and requires future research. 

In this study, the BAC violations in the simulated data were generated based on the knowledge inherent in LLMs, while the real data were obtained by reproducing attacks based on CVE reports. Strictly speaking, the real data cannot fully represent live network data. If live network data (regrettably, these data are unobtainable~\cite{el2024preserving}) were used for testing, the real data we established would also have limitations: training models with them might lead to insufficient generalization capabilities, and they do not cover undisclosed CVEs.

Furthermore, the target of BAC violation detection is the traffic generated by undiscovered CVEs. Therefore, we did not provide disclosed CVEs as knowledge to the large model, as this would undermine the credibility of our test results. Additionally, the excellent results obtained from cross-validation based on real data may be attributed to the fact that both the randomly split training and test sets contain traffic related to the same CVEs. As a result, the cross-validation results are highly ideal.

\section{Threat to Validity}

\textbf{Construct validity:} Since BAC violations are low-frequency events, we adopt $\text{F}_1$ and MCC, which have better discriminative ability for the performance of imbalanced learning.

\textbf{Internal validity:} The inaccessible of project specific traffic data with BAC violations precludes direct comparisons between simulated and real-world data during training. The public datasets employed are cross-project in nature, whereas our simulated data is purpose-built for the tested APIs. Consequently, the variable control in RQ2 has limitations, and our results only demonstrate that, relative to these exclusively available public datasets, our simulated dataset enables more effective model training for the specific task of BAC.


\textbf{Conclusion validity:} We analyzed the contributions of each component of the \BAC through ablation experiments and evaluated the contribution of each training feature via SHAP values, thereby demonstrating the value of the innovative points in this study.

\textbf{External validity:} Our test data (API-sets) are traffic data generated through real and valid accesses by multiple experts based on real reported CVEs, and we have conducted strict screening on the data. However, even so, these data were not directly obtained from real online operating environments, which may pose a threat to the validity of the research conclusions. Although our test set is already large enough (compared with existing research), the effectiveness of our method in more projects and its detection capability against attacks that may be suffered by CVEs not yet reported both require further validation.

\section{Conclusion}
\label{sec: conclusion}
To enhance the detection of BAC violations, we propose \BAC, a pioneering approach that utilizes composite API traffic generated by LLMs. \BAC introduces two core innovations: 1) we present a composite API traffic generation mechanism based on LLMs, which addresses the critical challenge of scarce real-world datasets that stem from the design constraints of RESTful architectures. By parsing API semantics and contextual correlations, it generates diverse simulated access requests with business logic, filling the gaps in coverage and semantic richness of data collection. 2) We trained a detection model with both sequential and static features; the multidimensional feature modeling can characterize the dynamic runtime information of multiple related traffic streams. By fusing dynamic interaction patterns with static semantic rules, the model’s accuracy in identifying context-dependent violations is improved. Our research demonstrates the significant value of LLM-generated simulated data in privacy-sensitive domains such as software security, where real-world datasets are often scarce or restricted.

\section*{Data Availability}

All source code and datasets used in this work are open source, and available at~\url{https://figshare.com/s/f1e41b9a2bcbb497765b}.

\footnotesize
\bibliographystyle{IEEEtranN}
\bibliography{AC}

\end{document}